\documentclass[aps,prb,twocolumn,preprintnumbers,amsmath,amssymb,superscriptaddress]{revtex4}%

\usepackage{graphicx}
\usepackage{dcolumn}
\usepackage{amsmath}
\usepackage{color}
\usepackage{hyperref}
\usepackage{bbold}
\usepackage{multirow}

\newcommand{\RN}[1]{\textup{\uppercase\expandafter{\romannumeral#1}}}%

\begin{document}

%\title{Magnetic response of 1313 structural polymorph of La$_3$Ni$_2$O$_7$ Ruddelsden-Popper nickelate}
\title{Magnetism of the alternating monolayer-trilayer phase of La$_3$Ni$_2$O$_7$}

\author{Rustem Khasanov}
 \email{rustem.khasanov@psi.ch}
 \affiliation{PSI Center for Neutron and Muon Sciences CNM, 5232 Villigen PSI, Switzerland}

\author{Thomas J. Hicken}
 \affiliation{PSI Center for Neutron and Muon Sciences CNM, 5232 Villigen PSI, Switzerland}

\author{Hubertus Luetkens}
 \affiliation{PSI Center for Neutron and Muon Sciences CNM, 5232 Villigen PSI, Switzerland}

\author{Zurab Guguchia}
 \affiliation{PSI Center for Neutron and Muon Sciences CNM, 5232 Villigen PSI, Switzerland}

\author{Dariusz J. Gawryluk}
 \affiliation{PSI Center for Neutron and Muon Sciences CNM, 5232 Villigen PSI, Switzerland}

\author{Vignesh Sundaramurthy}
 \affiliation{Max-Planck-Institute for Solid State Research, Heisenbergstra{\ss}e 1, 70569 Stuttgart, Germany}

\author{Abhi Suthar}
 \affiliation{Max-Planck-Institute for Solid State Research, Heisenbergstra{\ss}e 1, 70569 Stuttgart, Germany}

\author{Masahiko Isobe}
 \affiliation{Max-Planck-Institute for Solid State Research, Heisenbergstra{\ss}e 1, 70569 Stuttgart, Germany}

\author{Bernhard Keimer}
 \affiliation{Max-Planck-Institute for Solid State Research, Heisenbergstra{\ss}e 1, 70569 Stuttgart, Germany}

\author{Giniyat Khaliullin}
 \affiliation{Max-Planck-Institute for Solid State Research, Heisenbergstra{\ss}e 1, 70569 Stuttgart, Germany}

\author{Matthias Hepting}
 \affiliation{Max-Planck-Institute for Solid State Research, Heisenbergstra{\ss}e 1, 70569 Stuttgart, Germany}

\author{Pascal Puphal}
 \email{puphal@fkf.mpg.de}
 \affiliation{Max-Planck-Institute for Solid State Research, Heisenbergstra{\ss}e 1, 70569 Stuttgart, Germany}
 \affiliation{2.~Physikalisches Institut, Universit\"at Stuttgart, Pfaffenwaldring 57, 70569  Stuttgart, Germany}

\begin{abstract}

Understanding the magnetic ground state of Ruddlesden--Popper nickelates is crucial, as these materials
exhibit superconductivity under high pressure and host competing electronic orders
that may play a key role in the pairing mechanism. In this work, we investigate the magnetic
properties of the alternating monolayer-trilayer phase  of La$_3$Ni$_2$O$_7$ (1313-La$_3$Ni$_2$O$_7$) using muon-spin rotation/relaxation
($\mu$SR) under both ambient and hydrostatic pressure conditions.
The  monolayer-trilayer phase develops incommensurate magnetic order below approximately 150~K,
with a mean ordering temperature of $T_{\rm SDW} \simeq 123$~K and a transition width of
$\Delta T_{\rm SDW} \simeq 15$~K. The abrupt onset of the internal magnetic field indicates a
first-order-like transition. Hydrostatic pressure ($p$) suppresses the magnetic ordering temperature at a
rate of ${\rm d}T_{\rm SDW}/{\rm d}p \simeq -3.9$~K/GPa, demonstrating a progressive destabilization
of the ordered state.
By comparison with the bilayer 2222-La$_3$Ni$_2$O$_7$ and the trilayer 3333-La$_4$Ni$_3$O$_{10}$ systems, and within a unified phenomenological framework, systematic trends are identified linking the pressure dependence of $T_{\rm SDW}$, the (in)commensurability of the magnetic order, and the character of the magnetic transition. These trends consistently indicate a gradual reduction of electronic correlation strength from the bilayer to the monolayer–trilayer and trilayer nickelates. This hierarchy suggests that the higher superconducting transition temperature observed in the 2222 phase may be closely connected to its more strongly correlated electronic nature.
These results position the alternating monolayer-trilayer 1313-La$_3$Ni$_2$O$_7$ as an intermediate member linking the magnetic
behavior of the bilayer 2222-La$_3$Ni$_2$O$_7$ and the trilayer 3333-La$_4$Ni$_3$O$_{10}$ Ruddlesden--Popper
compounds.
Overall, our findings provide new insight into the interplay between structure, magnetism, and pressure in layered nickelates, contributing to a deeper understanding of the competing phases that may be relevant to the emergence of superconductivity.

\end{abstract}
\maketitle

\section{Introduction}

The discovery of pressure-induced superconductivity in the Ruddlesden--Popper (RP) nickelate La$_3$Ni$_2$O$_7$ has generated considerable interest in the condensed-matter physics community \cite{Sun_Nature_2023, Zhang_NatPhys_2024, Wang_Nature_2024, Liu_NatCom_2024, Zhang_JMST_2024, Li_ChinPhysLet_2024, Wang_ChinPhysLett_2024, Zhou_arxiv_2023}.
Initially, superconductivity was associated with the bilayer (2222) RP structure, characterized by two adjacent Ni--O planes separated by La--O blocking layers.
Subsequently, superconductivity was also reported in other RP structural variants, including the trilayer (3333) La$_4$Ni$_3$O$_{10}$,\cite{Li_CPL_La4Ni3O10_2024, Zhu_Nature_2024, Zhang_PRX_2025}
the alternating monolayer-trilayer (1313) La$_3$Ni$_2$O$_7$,\cite{Puphal_PRL_2024, Chen_JACS_2024, Wnag_InorgChem_2024, Abadi_PRL_2025, Huang_Arxiv_2025}
and an alternating monolayer-bilayer (1212) configuration such as La$_2$NiO$_4\cdot$La$_3$Ni$_2$O$_7$ \cite{Li_PRM_2024,Shi_NatPhys_2025}.

This structural diversity raises fundamental questions about the origin of superconductivity in RP nickelates.
Is superconductivity intrinsic to a particular layered architecture, or can it emerge from structural intergrowths, minority phases, or interfaces between distinct stacking units?
Clarifying these issues is essential for understanding the mechanisms responsible for superconductivity in these materials.

A powerful strategy to address these questions is to map and compare the pressure--temperature phase diagrams of different RP nickelate families.
For the bilayer La$_3$Ni$_2$O$_7$  and the trilayer La$_4$Ni$_3$O$_{10}$, such phase diagrams were established and consistently demonstrate the presence of density-wave orders preceding the appearance of superconductivity \cite{Sun_Nature_2023, Zhang_NatPhys_2024, Wang_Nature_2024, Li_ChinPhysLet_2024, Zhu_Nature_2024, Zhang_PRX_2025}.
At ambient pressure, both material types exhibit clear signatures of spin-density-wave (SDW) order \cite{Khasanov_NatPhys_2025, Chen_PRL_2024, Chen_arxiv_2024, Dan_SciBull_2025, Kakoi_JPSJ_2024, Chen_NatPhys_2024, Ren_CommPhys_2025, Gupta_NatCom_2025, Luo_CinPhysLett_2025, Khasanov_La4310_Arxiv_2025, Zhang_NatCom_2020}. The SDW state is commensurate in 2222 phase but becomes incommensurate in 3333 form, and the magnetic transition evolves from a second-order-like behavior in the bilayer system to a more abrupt, first-order-like transition in the trilayer compound \cite{Khasanov_NatPhys_2025, Khasanov_La4310_Arxiv_2025, Zhang_NatCom_2020, Cao_arxiv_2025, Chen_PRL_2024, Khasanov_LaPr327_Arxiv_2025, Plokhikh_Arxiv_2025, Chen_arxiv_2024}.
Charge-density-wave (CDW) order was directly observed in 3333-type La$_4$Ni$_3$O$_{10}$,\cite{Zhang_NatCom_2020}
whereas in 2222-La$_3$Ni$_2$O$_7$ the existence of a CDW instability is supported by optical pump--probe spectroscopy,\cite{Meng_NatCom_2024}
x-ray absorption near-edge spectroscopy,\cite{LiMintago_arxiv_2025}
and nuclear quadrupole resonance measurements \cite{Luo_CinPhysLett_2025}.

The pressure evolution of the SDW and CDW states further underscores the contrasting behavior of these RP systems.
In 3333 comounds, the CDW and SDW orders are intertwined, share a common transition temperature, and are monotonically suppressed under applied pressure \cite{Khasanov_La4310_Arxiv_2025}.
In 2222, by contrast, the CDW and SDW transitions occur at distinct temperatures and evolve largely independently:
the SDW transition temperature initially increases with pressure before being suppressed at higher pressures, eventually giving way to superconductivity \cite{Sun_Nature_2023, Zhang_NatPhys_2024, Wang_Nature_2024, Li_ChinPhysLet_2024, Zhu_Nature_2024, Zhang_PRX_2025}.

Despite these advances, the microscopic magnetic and density-wave states of the alternating monolayer-trilayer phase of La$_3$Ni$_2$O$_7$ (1313) remain poorly characterized.
Although angle-resolved photoemission spectroscopy data provide some indication of SDW order  at ambient pressure,\cite{Au-Yeung_arxiv_2025} the recently established pressure--temperature phase diagram of 1313-La$_3$Ni$_2$O$_7$ is based primarily on transport signatures of superconductivity,\cite{Huang_Arxiv_2025}
leaving open several key questions regarding the nature and evolution of ordered states at ambient and moderate pressures:

\begin{enumerate}
    \item What types of density-wave states are present in the 1313 phase, and can they serve as precursors to superconductivity?
    \item Do these ordered states behave similarly to those observed in the 2222 or the 3333 phases, or do they exhibit qualitatively different evolution with pressure?
    \item How do these competing orders connect to, or potentially interact with superconductivity observed in the 1313 phase under high pressure \cite{Huang_Arxiv_2025}?
\end{enumerate}

In this work, we attempt to address these questions by performing muon-spin rotation/relaxation ($\mu$SR) experiments on samples predominantly composed of the 1313 phase, both at ambient and under hydrostatic pressure.
Our measurements reveal that the magnetic order in 1313 phase is incommensurate and that the transition into the magnetic state is first-order-like, suggesting the presence of an intertwinned CDW instability, analogous to the mechanism proposed for the 3333-La$_4$Ni$_3$O$_{10}$ \cite{Zhang_NatCom_2020}.
These findings establish the alternating monolayer-trilayer 1313-La$_3$Ni$_2$O$_7$ as an intermediate member linking the magnetic behavior of the bilayer 2222-La$_3$Ni$_2$O$_7$ and the trilayer 3333-La$_4$Ni$_3$O$_{10}$ RP nickelates.

\section{Experimental Details}

La$_3$Ni$_2$O$_7$ single crystals were grown using the optical floating-zone technique under high oxygen pressure. Details of the crystal growth procedure are provided in Ref.~\onlinecite{Puphal_PRL_2024}. All crystals investigated in this work were annealed in 600~bar O$_2$ at 600$^{\circ}$C for 7~days, followed by quenching to room temperature.

Powder x-ray diffraction (PXRD) measurements were performed on finely ground crystal pieces to check for impurity phases and to determine the crystal structure and lattice parameters. The PXRD patterns were recorded at room temperature using a Rigaku Miniflex diffractometer in Bragg--Brentano geometry with Cu $K_{\alpha}$ radiation and a Ni filter. Rietveld refinements were carried out using the TOPAS~V6 software package.

Magnetic susceptibility measurements were performed using a vibrating sample magnetometer (MPMS-3, Quantum Design) in magnetic fields up to 7~T and temperatures between 1.8 and 350~K.

Polarized optical microscopy experiments were conducted using Leitz Orthoplan microscope 933928, equipped with crossed polarizers. Prior to the measurements, several crystals were polished using diamond slurry with particle sizes down to 0.1~$\mu$m to obtain optically flat surfaces suitable for domain imaging.

Raman spectroscopy measurements were carried out using a HORIBA Jobin Yvon LabRAM HR800 spectrometer equipped with a He--Ne laser (wavelength 632.8~nm). A 50$\times$ ultra-long-working-distance objective was used, with a laser power of approximately 1~mW at the sample position. The Raman spectra presented in this work were divided by the Bose--Einstein factor.

Ambient- and high-pressure muon-spin rotation/relaxation ($\mu$SR) experiments were performed at the $\pi$M3 and $\mu$E1 beamlines of the Paul Scherrer Institute (PSI), Villigen, Switzerland, using the GPS (General Purpose Surface) spectrometer~\cite{Amato_RSI_2017} and the GPD (General Purpose Decay) spectrometer~\cite{Khasanov_HPR_2016, Khasanov_JAP_2022}. Quasi-hydrostatic pressures up to $\simeq 2.3$~GPa were generated using double-wall piston-cylinder clamp cells fabricated from nonmagnetic MP35N alloy~\cite{Khasanov_HPR_2016}. $\mu$SR measurements were performed in two geometries: zero-field (ZF) $\mu$SR, conducted in the absence of an external magnetic field, and  weak-transverse-field (WTF) $\mu$SR, with a small magnetic field applied perpendicular to the initial muon-spin polarization. Descriptions of the $\mu$SR technique applied to the study of magnetic and superconducting materials can be found in several review articles and textbooks, see {\it e.g.} Refs.~\onlinecite{Amato-Morenzoni_book_2024, Schenk_book_1995, Yaouanc_book_2011, Blundell_book_2022, Blundell_AnnRev_2025}. The $\mu$SR data were analyzed using the \textsc{MUSRFIT} software package~\cite{MUSRFIT}.

\section{Results }

\subsection{Sample Characterisation}\label{sec:optical-characterzation}

\begin{figure}[htb]
\includegraphics[width=0.9\linewidth]{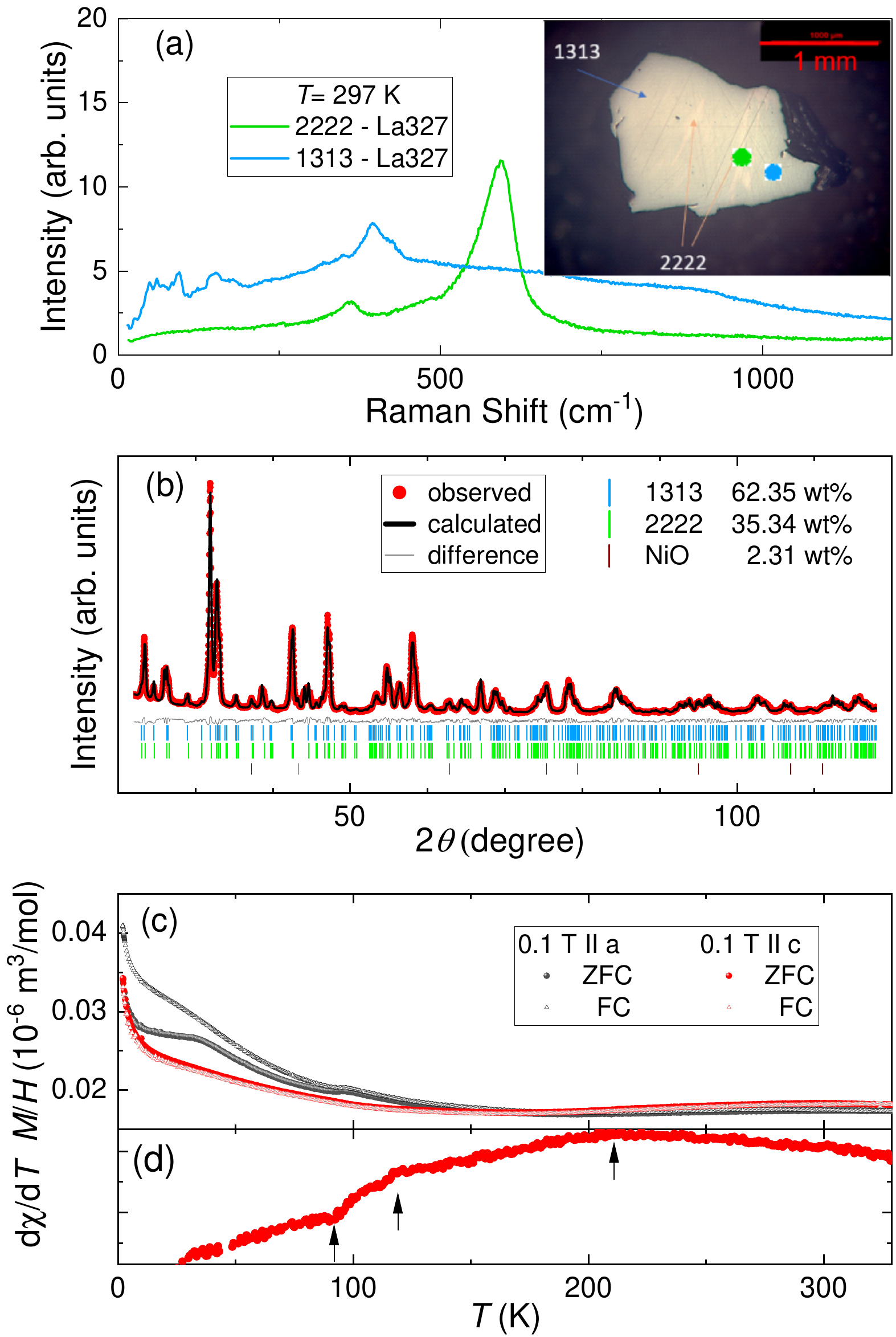}
\caption{
(a) Raman spectra from selected positions on a La$_3$Ni$_2$O$_7$ single crystal.
The green and blue curves correspond to responses obtained from regions enriched in the
2222 and 1313 structural phases of La$_3$Ni$_2$O$_7$, respectively.
The inset shows a polarized optical microscopy image of the same crystal;
the green and blue markers indicate the locations where the corresponding Raman spectra were collected.
(b) Powder x-ray diffraction pattern of a crushed single crystal together with a Rietveld refinement.
The solid black line represents the calculated intensity from the refinement,
the solid gray line shows the difference between the measured and calculated intensities,
and the vertical blue/green bars mark the calculated Bragg peak positions.
(c) Magnetic susceptibility of the La$_3$Ni$_2$O$_7$ single crystal shown in panel (a),
measured in an applied field of 0.1~T along the $c$ axis (red) and within the $ab$ plane (black).
Filled spheres denote zero-field-cooled (ZFC) data,
while open triangles represent field-cooled (FC) measurements.
(d) Temperature derivative of the susceptibility curve measured with the field applied along the $c$ axis
[red symbols in panel (c)].
Arrows mark the anomalies associated with magnetic transitions.
}
 \label{fig:Polarised-microscopy}
\end{figure}

La$_3$Ni$_2$O$_7$ grows only within a narrow range of oxygen partial pressures and typically forms as a polymorphic mixture containing different stacking variants. These include the bilayer 2222 and the monolayer-trilayer 1313 phases, as described in detail in the Supplementary Information. Several selected single crystals were polished and examined using polarized optical microscopy, Raman spectroscopy and magnetic susceptibility measuremements. Although individual crystals can appear relatively phase-pure, combined analysis reveals that the La$_3$Ni$_2$O$_7$ boule contains multiple structural phases.

A representative polarized optical microscopy image of a selected single crystal is shown in the inset of Fig.~\ref{fig:Polarised-microscopy}(a). Even in this nominally phase-pure sample, polarization contrast reveals minority inclusions -- such as the 2222 variant -- embedded within a matrix dominated by the 1313 phase. Both phases can be unambiguously distinguished through their characteristic Raman signatures, as presented in Fig.~\ref{fig:Polarised-microscopy}(a) and Ref.~\onlinecite{Sundaramurthy_arxiv_2025}.

Phase compositions within individual crystals and across the bulk were further evaluated by crushing selected crystals and performing powder x-ray diffraction, as shown in Fig.~\ref{fig:Polarised-microscopy}(b). Rietveld refinements yield lattice parameters of
1313 phase: $a = 5.4378$~\AA, $b = 5.4707$~\AA, $c = 20.2835$~\AA; and
2222 phase: $a = 5.3956$~\AA, $b = 5.4516$~\AA, $c = 20.5207$~\AA,
with refinement statistics $R_{\mathrm{exp}} = 1.37$, $R_{\mathrm{wp}} = 4.67$, and $\mathrm{GOF} = 3.40$.
The elongated $c-$axis of the 2222 phase compared with the 1313 variant is consistent with literature values.

Magnetic susceptibility measurements of the crystal shown in Fig.~\ref{fig:Polarised-microscopy}(a) reveal several magnetic anomalies, leading to the complex temperature dependence displayed in Fig.~\ref{fig:Polarised-microscopy}(c). The overall shape of the magnetization curves is similar to that reported in Ref.~\onlinecite{Puphal_PRL_2024} for a crystal originating from the same growth batch.
The derivative of the magnetization measured with the field applied along the $c-$axis,
Fig.~\ref{fig:Polarised-microscopy}(d), identifies three distinct magnetic transitions:
one near 200~K, a second around 120~K, and a third near 100~K.
The low-temperature transition correlates with the bifurcation between field-cooled and zero-field-cooled magnetization measured along the $a-$axis.
Although PXRD shows no explicit evidence for La$_2$NiO$_4$, the weak anomaly near 200~K may indicate a small fraction of this phase.

\begin{figure}[t]
		\includegraphics[width=0.9\columnwidth]{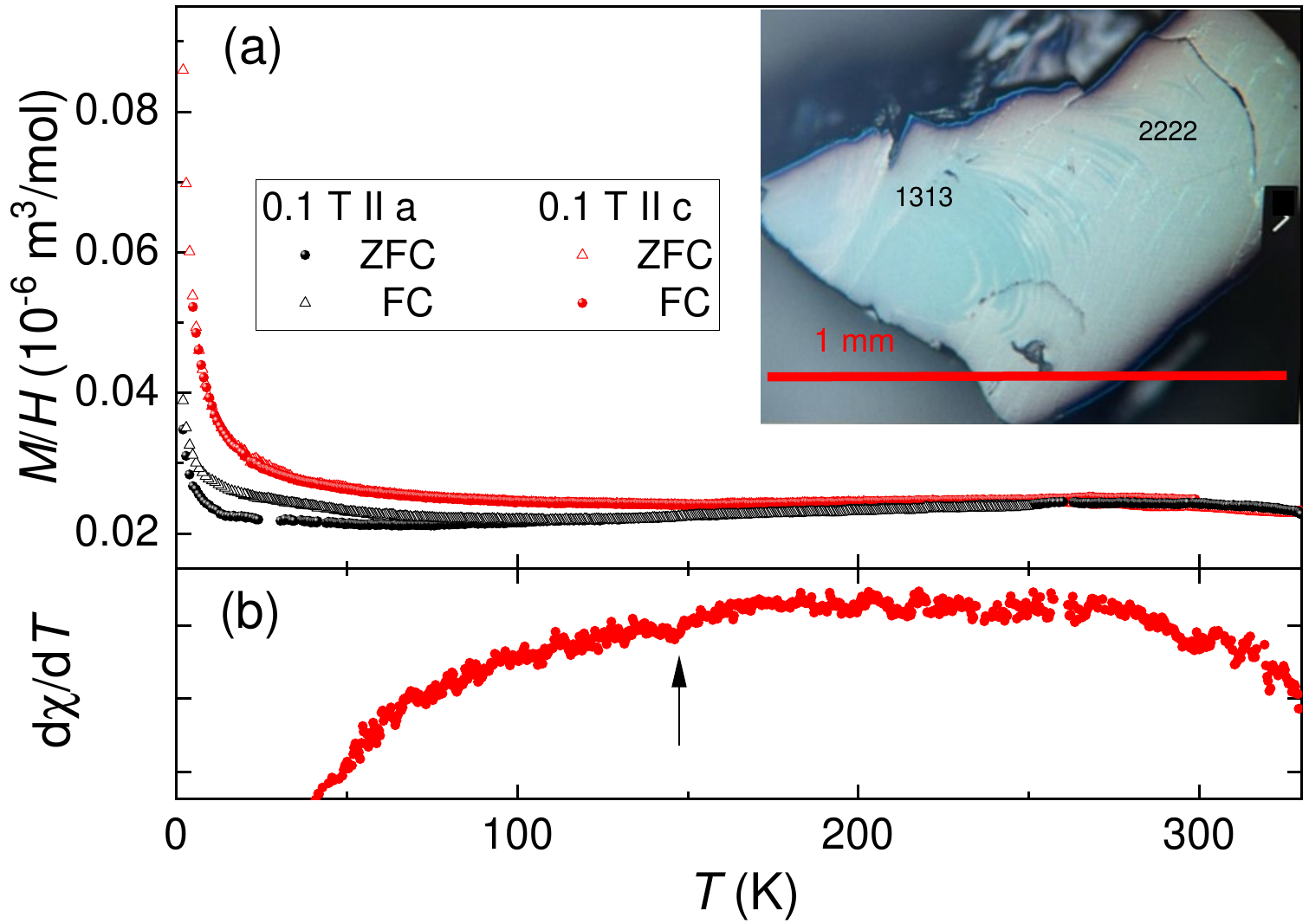}
	\caption{
(a) Magnetic susceptibility of another La$_3$Ni$_2$O$_7$ single crystal measured in an applied field of 0.1~T
along the $c$ axis (red) and within the $ab$ plane (black). Filled spheres denote zero-field-cooled (ZFC) data,
while open triangles represent field-cooled (FC) measurements. The inset shows a polarized optical microscopy
image of this crystal.
(b) Temperature derivative of the susceptibility curve measured with the field applied along the $c$ axis
[red symbols in panel (a)].
}
	\label{2222squid}
\end{figure}

The magnetization curves of another single crystal are shown in Fig.~\ref{2222squid}(a).
A region predominantly composed of the 2222 phase was identified in this sample, as indicated in the inset of Fig.~\ref{2222squid}(a).
The corresponding magnetization data exhibit a much simpler temperature dependence than that of the 1313 variant.
The derivative of the magnetization measured along the $c-$axis, Fig.~\ref{2222squid}(b), displays a single, well-defined transition around 150~K.
A gradual change in the susceptibility slope at higher temperatures, visible in both samples, is characteristic of low-dimensional quantum magnets and does not correspond to a genuine phase transition.

It should be noted, that individually selected single crystals are relatively phase-pure and typically contain only mixtures of the 2222- and 1313-La$_3$Ni$_2$O$_7$ phases. Importantly, the crystals used in earlier studies reported in Refs.~\onlinecite{Puphal_PRL_2024} and \onlinecite{Abadi_PRL_2025} were taken from the same growth batch as the material used for the $\mu$SR experiments.
However, $\mu$SR measurements require a substantially larger sample mass, which necessitated crushing the entire growth boule. As detailed in the Supplementary Information, the resulting bulk powder contains additional impurity phases, including neighboring Ruddlesden--Popper members La$_2$NiO$_4$ and La$_4$Ni$_3$O$_{10}$, which are either absent or strongly suppressed in individually selected crystals but become apparent when averaging over the whole boule.

\subsection{Muon-Spin Rotation/Relaxation Experiments}

\subsubsection{Weak-Transverse Field $\mu$SR Experiments}\label{subsec:WTF}

Experiments in a weak transverse field, applied perpendicular to the initial muon-spin polarization, were performed to probe the temperature evolution of the magnetic volume fraction. Spins of muons stopped in a paramagnetic (nonmagnetic) environment precess coherently in the externally applied field, producing a well-defined oscillation of the asymmetry signal. In contrast, muons stopping in magnetically ordered regions experience an internal magnetic field that is the vector sum of the external field and the local static fields from ordered moments. Because these internal fields are typically large and broadly distributed, the corresponding muon spins depolarize rapidly, effectively suppressing the observable oscillating WTF signal from the magnetic fraction.
Thus, the reduction of the oscillation amplitude directly reflects the growth of magnetically ordered regions in the sample. For further details of WTF-$\mu$SR experiments, see Ref.~\onlinecite{Amato-Morenzoni_book_2024}.

Two types of WTF-$\mu$SR experiments were conducted: measurements at ambient pressure and measurements under hydrostatic-pressure conditions up to a maximum pressure of $\simeq 2.3$~GPa.

\begin{figure}[htb]
    \centering
    \includegraphics[width=0.8\linewidth]{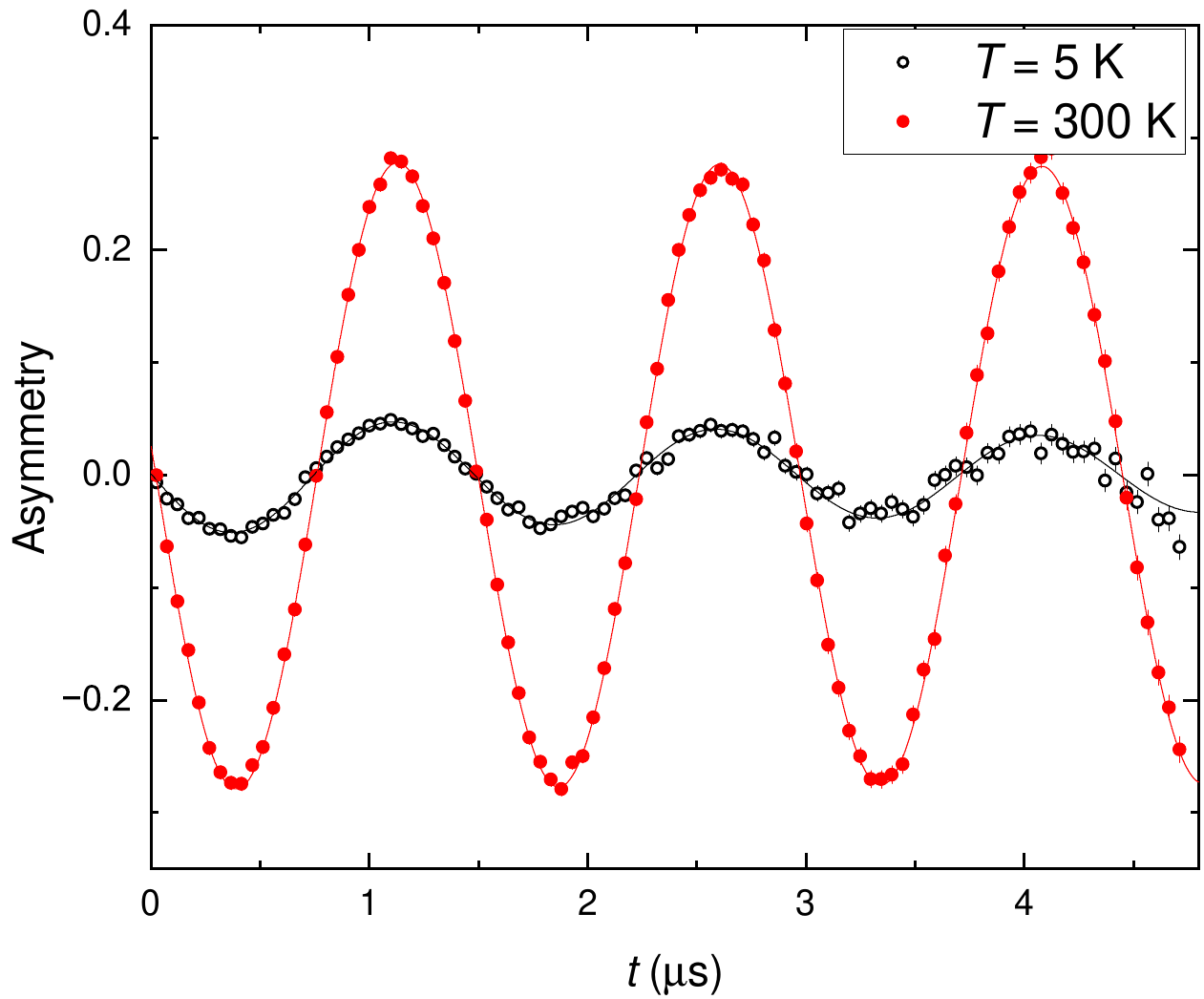}
    \caption{Weak transverse field (WTF) $\mu$SR time spectra collected at $T = 5$~K and 300~K. Solid lines represent fits using Eq.~\ref{eq:WTF}.}
    \label{fig:WTF_time-spectra}
\end{figure}

\vspace{0.5cm}
\noindent {\it WTF-$\mu$SR at ambient pressure.}
Figure~\ref{fig:WTF_time-spectra} shows the $\mu$SR time spectra collected at $T = 5$~K and 300~K in a weak transverse magnetic field $B_{\rm WTF} = 5$~mT applied perpendicular to the initial muon-spin polarization.
At $T = 300$~K, a clear oscillatory signal corresponding to muon-spin precession in $B_{\rm WTF}$ is observed. The initial asymmetry $A_0 \simeq 0.28$ matches the maximum value expected for the GPS spectrometer in transverse field mode,\cite{Amato_RSI_2017} indicating that the La$_3$Ni$_2$O$_7$ sample is fully paramagnetic ({\it i.e.}, nonmagnetic) at room temperature. Upon cooling to 5~K, the oscillatory component is significantly reduced, suggesting the appearance of static internal magnetic fields and partial magnetic ordering. However, the oscillations do not vanish completely, implying that a portion of the sample volume (approximately 20\%) remains nonmagnetic down to the lowest temperature measured.  This value is notably higher than the typical background of the GPS instrument -- $\simeq 3$–5\%, Ref.~\onlinecite{Amato_RSI_2017}.

The WTF asymmetry signal was modeled using the following expression:
\begin{equation}
A(t) = A_0 \; (1 - f_{\rm m})\; e^{-\sigma_{\rm WTF}^2 t^2 / 2}\; \cos(\gamma_\mu B_{\rm WTF} t + \phi),
\label{eq:WTF}
\end{equation}
which describes muon-spin precession in the nonmagnetic portion of the sample. Here, $f_{\rm m}$ is the magnetic volume fraction, $\phi$ is the initial phase of the muon-spin ensemble, $B_{\rm WTF} = 5$~mT is the applied field, $\gamma_\mu=135.538817$~Mhz/T is the muon gyromagnetic ratio, and $\sigma_{\rm WTF}$ is the Gaussian relaxation rate representing field inhomogeneity within the nonmagnetic regions.

\begin{figure}[htb]
    \centering
    \includegraphics[width=0.8\linewidth]{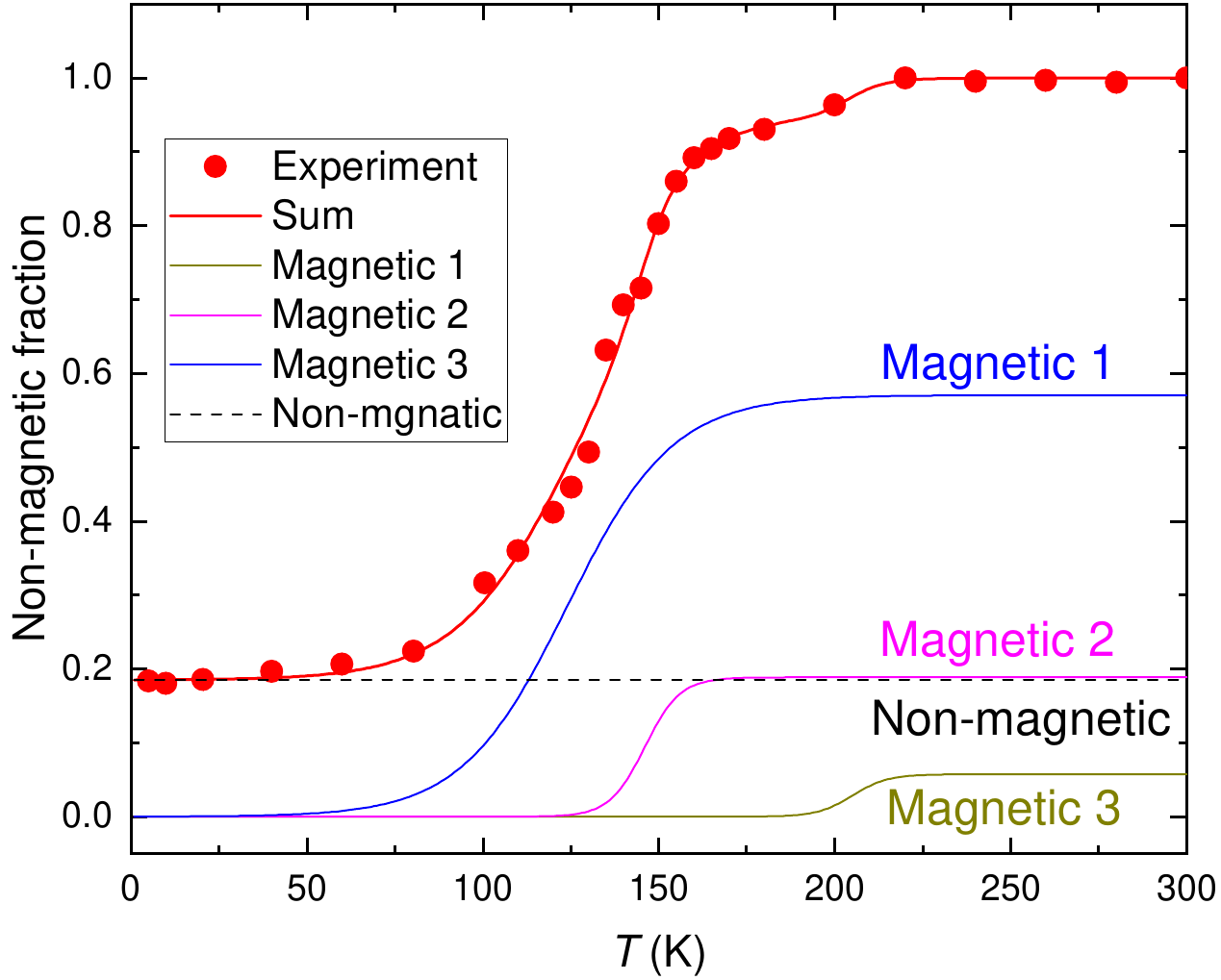}
    \caption{Temperature dependence of the nonmagnetic volume fraction $1 - f_{\rm m}$ extracted from WTF $\mu$SR measurements. The solid line represents a fit using Eq.~\ref{eq:frac4}, assuming three magnetic and one nonmagnetic component. The individual contributions to $1 - f_{\rm m}(T)$, as well as their total sum, are represented by the solid and dashed lines. See text for details.}
    \label{fig:WTF_nonmagnetic-fraction}
\end{figure}

Figure~\ref{fig:WTF_nonmagnetic-fraction} displays the temperature dependence of the nonmagnetic volume fraction $1 - f_{\rm m}$. The transition to the magnetic state,  which causes reduction of $1-f_{\rm m}$ with decreasing temperature, appears non-monotonic, suggesting the presence of multiple magnetic phases. The full temperature dependence was modeled assuming three magnetic components ($f_{{\rm m},i}$, $i = 1 \ldots 3$) and one nonmagnetic component ($f_{\rm nm}$):
\begin{equation}
1 - f_{\rm m}(T) = \sum_{i=1}^{3} \frac{f_{{\rm m},i}}{1 + \exp\left[\frac{T - T_{{\rm m},i}}{\Delta T_{{\rm m},i}}\right]} + f_{\rm nm},
\label{eq:frac4}
\end{equation}
where $T_{{\rm m},i}$ and $\Delta T_{{\rm m},i}$ are the midpoint and width of the $i$-th magnetic transition, and $f_{{\rm m},i}$ and $f_{\rm nm}$ are the respective volume fractions.

Analysis of the $1 - f_{\rm m}(T)$ data reveals that approximately 18\% of the sample volume remains nonmagnetic over the entire temperature range studied. The remaining portion exhibits three distinct magnetic transitions at $T_{\rm m} \simeq 124$~K, 145~K, and 200~K, with corresponding volume fractions of approximately 60\%, 15\%, and 5\%, respectively. The individual contributions to $1 - f_{\rm m}(T)$, as well as their total sum, are shown by the solid and dashed lines in Fig.~\ref{fig:WTF_nonmagnetic-fraction}.

\begin{figure}[htb]
\includegraphics[width=0.8\linewidth]{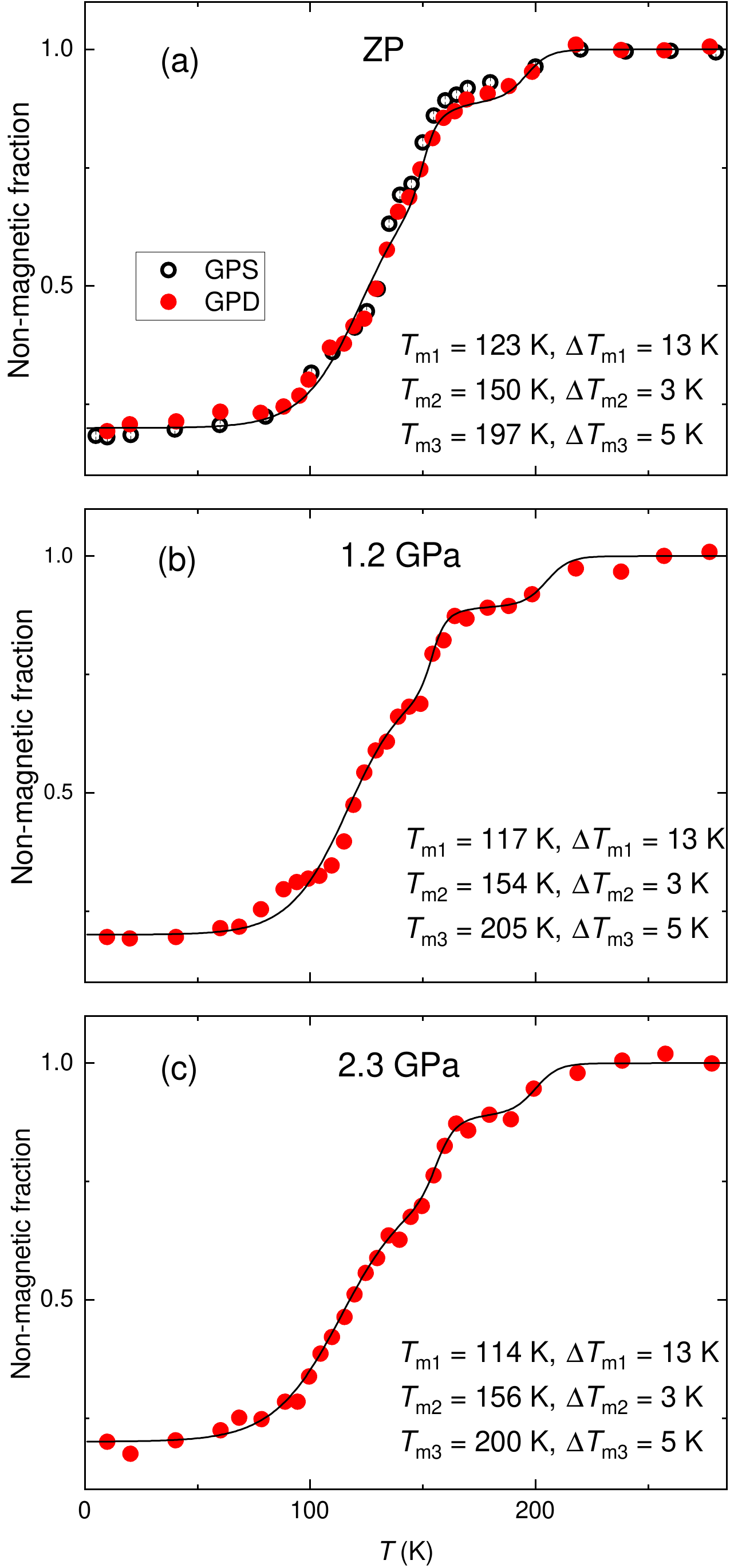}
\caption{(a) Temperature dependence of the nonmagnetic volume fraction $1 - f_{\rm m}$ at $p = 0.0$~GPa. Open symbols correspond to data from the low-background GPS instrument, see Fig.~\ref{fig:WTF_nonmagnetic-fraction}. (b) and (c) Temperature dependence of $1 - f_{\rm m}$ measured at $p = 1.2$ and 2.3~GPa. Solid lines are fits to Eq.~\ref{eq:frac4} with parameters shown in each panel.}
\label{fig:pressure_WTF}
\end{figure}

\vspace{0.5cm}
\noindent{\it WTF-$\mu$SR Under Pressure.}
Figure~\ref{fig:pressure_WTF} shows the temperature dependence of the nonmagnetic volume fraction $1 - f_{\rm m}$ at $p = 0.0$, 1.2, and 2.3~GPa. Panel (a) also compares ambient-pressure WTF results obtained on the GPS instrument (low-background setup) with those measured using the GPD spectrometer (sample inside pressure cell). The two data sets at ambient pressure agree well, confirming measurement consistency between setups.

The solid lines in Fig.~\ref{fig:pressure_WTF} represent fits of $1 - f_{\rm m}(T)$ using Eq.~\ref{eq:frac4}. All pressure data sets were fitted simultaneously under the assumption that the magnetic volume fractions and transition widths of each phase remain pressure-independent. The extracted magnetic ordering temperatures $T_{{\rm m},i}$ and transition widths $\Delta T_{{\rm m},i}$ are indicated in each panel. The volume fractions of the magnetic components were found to be $f_{\rm m,1} \simeq 0.54$, $f_{\rm m,2} \simeq 0.16$, and $f_{\rm m,3} \simeq 0.10$, with a nonmagnetic component of $f_{\rm nm} \simeq 0.20$. These values are similar, but not identical, to those obtained at ambient pressure (Fig.~\ref{fig:WTF_nonmagnetic-fraction}). The slight differences may be attributed to the larger sample volume and, therefore to slightly different material, used in $\mu$SR under pressure experiments.

\begin{figure}[htb]
\includegraphics[width=0.8\linewidth]{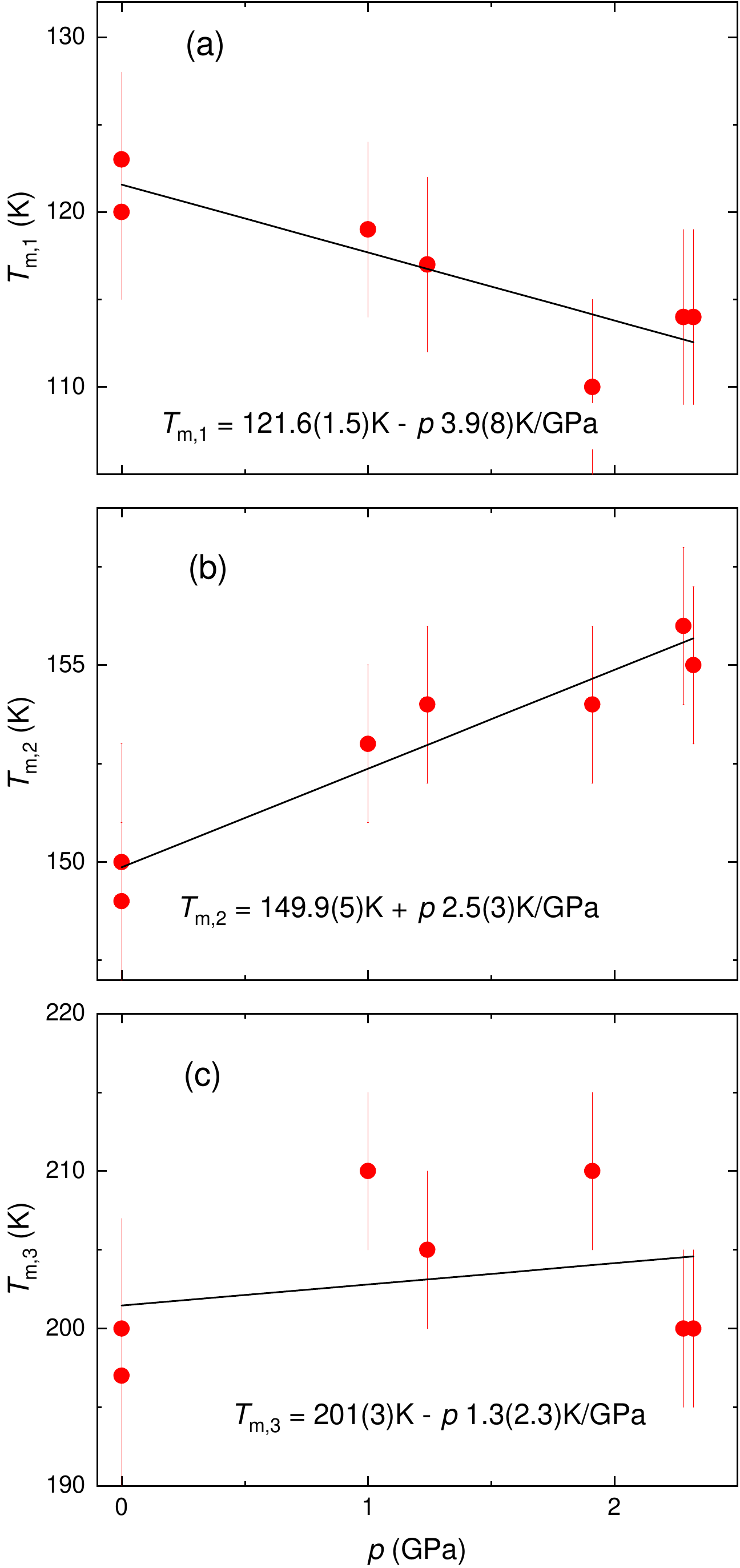}
\caption{(a) Pressure dependence of the magnetic ordering temperature $T_{\rm m}$ for the 1313 phase. (b) Pressure dependence of $T_{\rm m}$ for the 2222 phase. (c) Pressure dependence of $T_{\rm m}$ for the high-temperature magnetic phase (presumably La$_2$NiO$_4$). Solid lines represent linear fits to the data.
}
\label{fig:pressure_Tm}
\end{figure}

The pressure dependence of the magnetic ordering temperatures is summarized in Fig.~\ref{fig:pressure_Tm}. The error bars represent the widths of the corresponding transitions. The magnetic fractions exhibit distinct responses to applied pressure. The transition temperature of the dominant magnetic component [approximately 55\% of the volume; panel~(a)] decreases with pressure, whereas that of the second magnetic component [approximately 15\%; panel~(b)] increases. Notably, this opposite pressure evolution of the first and second magnetic components is reflected in the more pronounced three-step structure in $1 - f_{\rm m}(T)$ at higher pressures, as shown in Figs.~\ref{fig:pressure_WTF}~(b) and (c). The third magnetic component, with an ordering temperature near 200~K, Fig.~\ref{fig:pressure_Tm}~(c), remains nearly pressure independent.

\begin{figure*}[htb]
\includegraphics[width=1.0\linewidth]{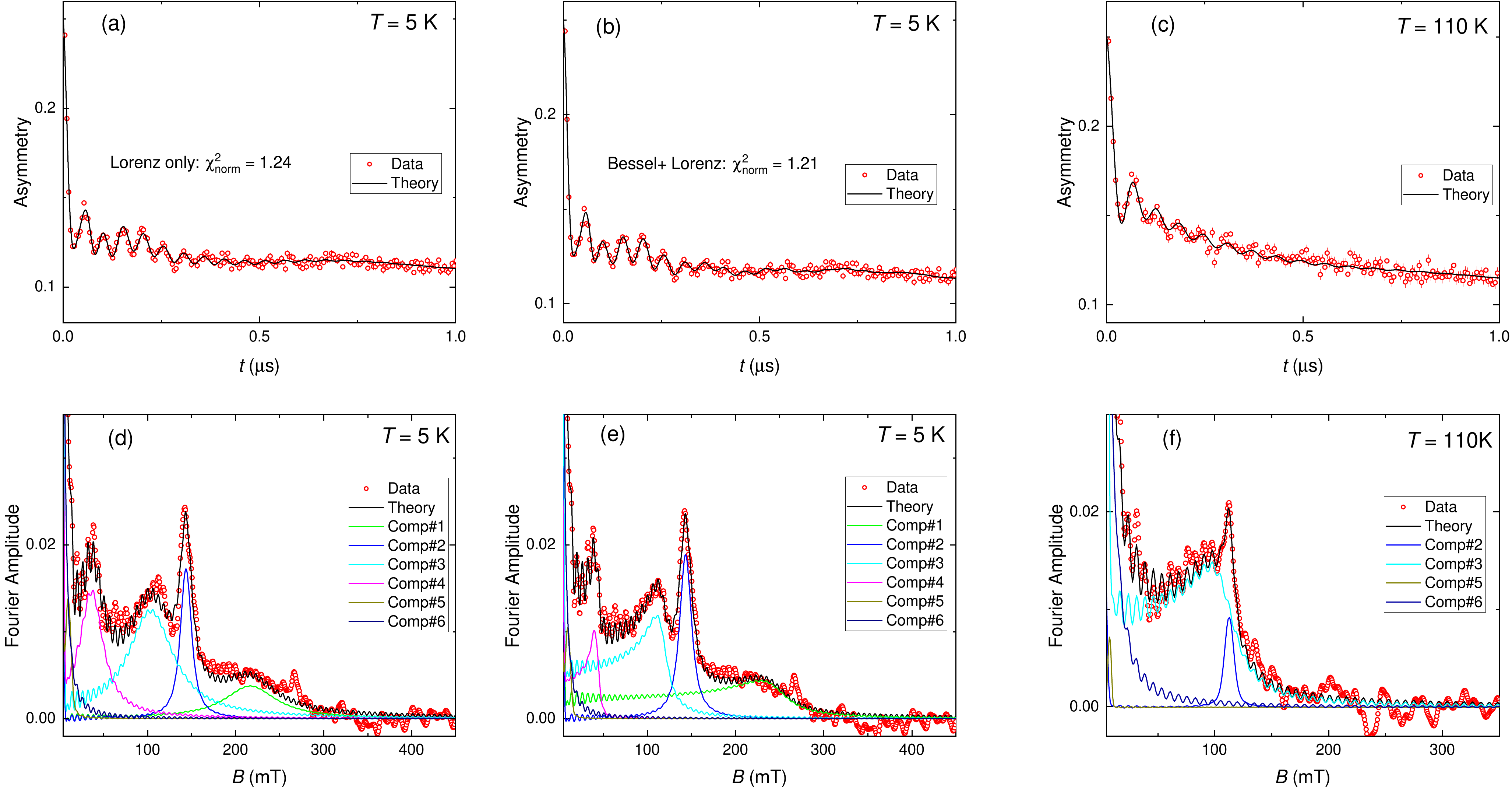}
\caption{(a)--(c) ZF-$\mu$SR time spectra measured at $T = 5$~K [panels (a) and (b)] and $T = 110$~K [panel (c)]. Solid lines represent fits using Eq.~\ref{eq:ZF_Lorenzian-Bessel} with different numbers of internal field components: 5 in panels (a) and (b), and 3 in panel (c). (d)--(e) Fourier transforms of the data in panels (a)--(c). Individual magnetic components are indicated by solid lines with different colors. See text for details.}
\label{fig:ZF-signals}
\end{figure*}

The pressure dependencies of the transition temperatures were fitted linearly, yielding the following results:
\begin{align*}
T_{\rm m,1}(p) &= 121.6(1.3){\rm ~K} - p\cdot 3.9(8){\rm ~K/GPa}, \\
T_{\rm m,2}(p) &= 149.9(5){\rm ~K} + p\cdot 2.5(3){\rm ~K/GPa}, \\
T_{\rm m,3}(p) &= 201(3){\rm ~K} + p\cdot 1.3(2.3){\rm ~K/GPa}.
\end{align*}
These results reveal a competition between different magnetic phases, with one component being suppressed, a second one enhanced, and a third remaining nearly stable under pressure. This behavior may reflect underlying structural differences and distinct magnetic responses among the phases.

\subsubsection{ZF-$\mu$SR Experiments}\label{subsec:ambient-pressure-ZF}

In zero-field (ZF) $\mu$SR experiments, no external magnetic field is applied, allowing the muon spins to respond solely to the internal magnetic fields arising from static or fluctuating electronic moments in the sample. Similar to WTF-$\mu$SR, ZF measurements provide information on the magnetic volume fraction. In addition, ZF-$\mu$SR offers direct access to the magnitude of the internal field(s) at the muon stopping site, which are proportional to the size of the ordered magnetic moments. This makes ZF-$\mu$SR particularly powerful for characterizing the nature and strength of magnetic order \cite{Amato-Morenzoni_book_2024}. The ZF-$\mu$SR measurements in this study were performed only at ambient pressure.

Figure~\ref{fig:ZF-signals} shows the zero-field  $\mu$SR time spectra and their corresponding Fourier transforms collected at $T = 5$~K [panels (a), (b), (d), and (e)] and $T = 110$~K [panels (c) and (f)]. The oscillatory behavior indicates the presence of internal magnetic fields at the muon stopping sites, originating from static magnetic ordering. The Fourier transforms reveal the magnetic field distribution, which consists of both symmetric and asymmetric peaks.

The ZF-$\mu$SR spectra were analyzed using the following functional form:
\begin{eqnarray}
A(t) &=& A_0\; f_{\rm m}\; \left[\frac{2}{3} \sum_{i=1}^{n} f_i e^{-\lambda_{T_i} t} F(\gamma_\mu B_i t) + \frac{1}{3} e^{-\lambda_L t} \right] \nonumber \\
&& + A_0 \; (1 - f_{\rm m})\; \mathrm{GKT}(t),
\label{eq:ZF_Lorenzian-Bessel}
\end{eqnarray}
where the first term represents the magnetic contribution and the second term corresponds to the nonmagnetic one, with relative weights $f_{\rm m}$ and $1 - f_{\rm m}$, respectively. The function $F(\gamma_\mu B_i t)$ describes muon spin precession in the internal field $B_i$ and may take the form of either a cosine function [$\cos(\gamma_\mu B_i t)$] or a zeroth-order Bessel function [$J_0(\gamma_\mu B_i t)$]. The coefficients $f_i$ are the volume fractions of the individual oscillatory components (with $\sum f_i = f_{\rm m}$), $n$ is the number of components, $\lambda$ denotes exponential relaxation rates, and $\mathrm{GKT}(t)$ is the Gaussian Kubo-Toyabe function accounting for nuclear magnetic contributions in the paramagnetic state \cite{Amato-Morenzoni_book_2024, Schenk_book_1995, Yaouanc_book_2011, Blundell_book_2022}.
The prefactors $2/3$ and $1/3$ arise from powder averaging, where $2/3$ of muon spins experience transverse (T) fields and $1/3$ remain longitudinally (L) aligned with the internal field direction \cite{Amato-Morenzoni_book_2024, Schenk_book_1995, Yaouanc_book_2011, Blundell_book_2022}.

The analysis reveals that the low-temperature response consists of six components: five internal field components (denoted as \#1 to \#5) and one fast-relaxing component (\#6, with $B_{\rm int,6} = 0$); see Figs.~\ref{fig:ZF-signals}~(a), (b), (d), and (e). Two fitting models were considered: one in which all internal fields are described by symmetric Lorentzian lineshapes [panels (a) and (d)], and another in which two fields are Lorentzian and three follow an Overhauser-type distribution, modeled using a zeroth-order Bessel function in the time domain [panels (b) and (e)] \cite{Overhauser_JPhysChemSolids_1960, Schenck_PRB_2001, Amato_PRB_2014, Khasanov_PRB_MnP_2016}. The latter model provides a better fit, as indicated by a smaller reduced chi-squared value: $\chi^2_{\rm n} \simeq 1.21$ compared to 1.24 for the all-Lorentzian case.

\begin{figure}[htb]
\includegraphics[width=0.8\linewidth]{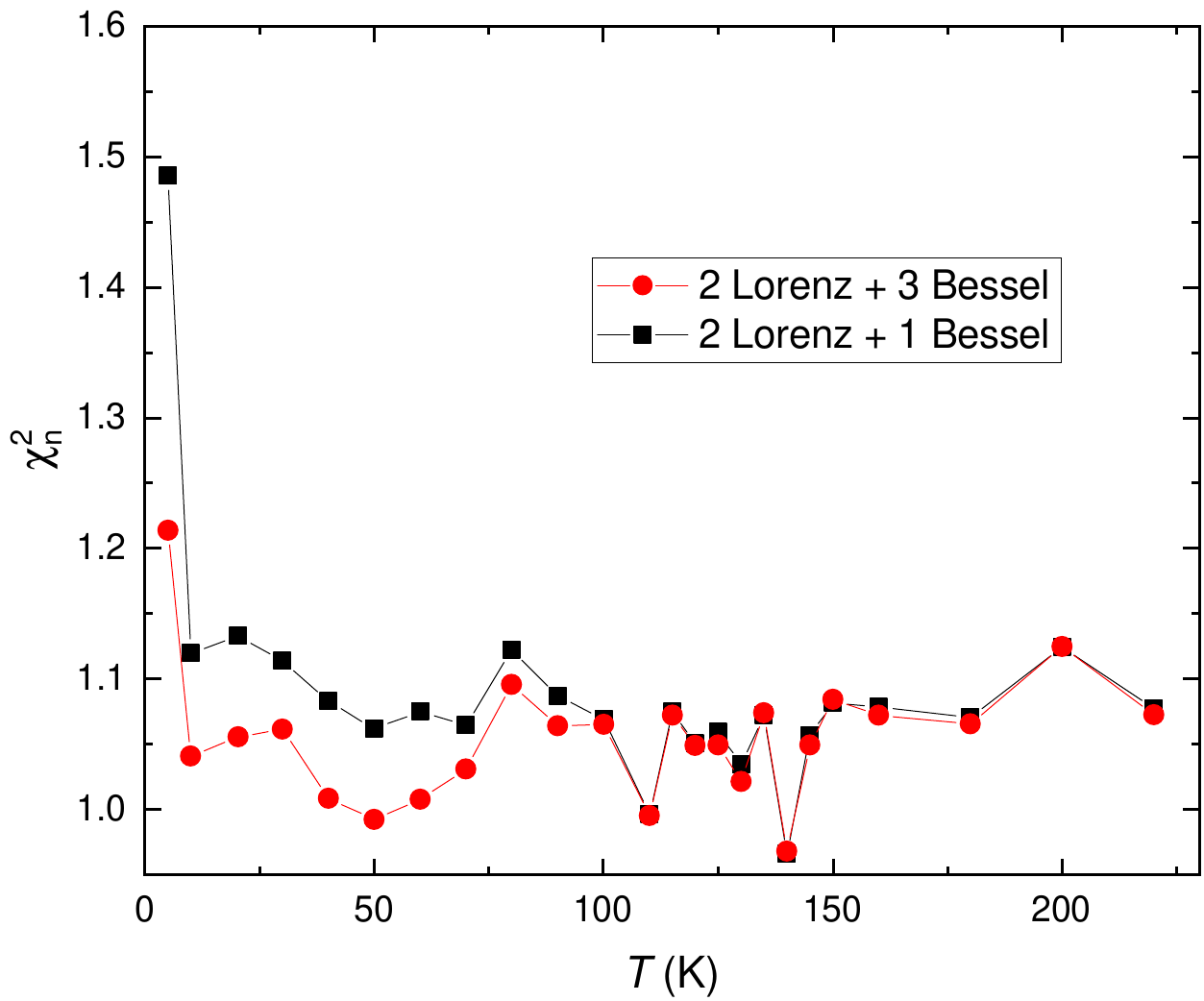}
\caption{Temperature dependence of the reduced chi-squared $\chi^2_{\rm n}$ for models with five and three internal field components. The divergence of $\chi^2_{\rm n}$ values below $\sim 100$~K indicates a spin-reorientation transition in the 1313-La$_3$Ni$_2$O$_7$ phase.}
\label{fig:chi2_5peak-3peak}
\end{figure}

At higher temperatures, the number of resolved internal fields reduces to three, Figs.~\ref{fig:ZF-signals}~(c) and (f), indicating a change in the magnetic structure of the studied sample  with temperature. Only one of the remaining internal field components retains an asymmetric lineshape, while the components \#1 and \#4 observed at low temperatures are no longer visible. To determine the temperature at which the magnetic field distribution changes from the five- to three-peak structure, Fig.~\ref{fig:chi2_5peak-3peak} shows the temperature evolution of the reduced chi-squared parameter for both fitting scenarios. The $\chi^2_{\rm n}$ values converge above 100~K and begin to diverge below this temperature, indicating the onset of another magnetic (probably spin-reorientation) transition around $T \simeq 100$~K.

\begin{figure}[htb]
\includegraphics[width=0.8\linewidth]{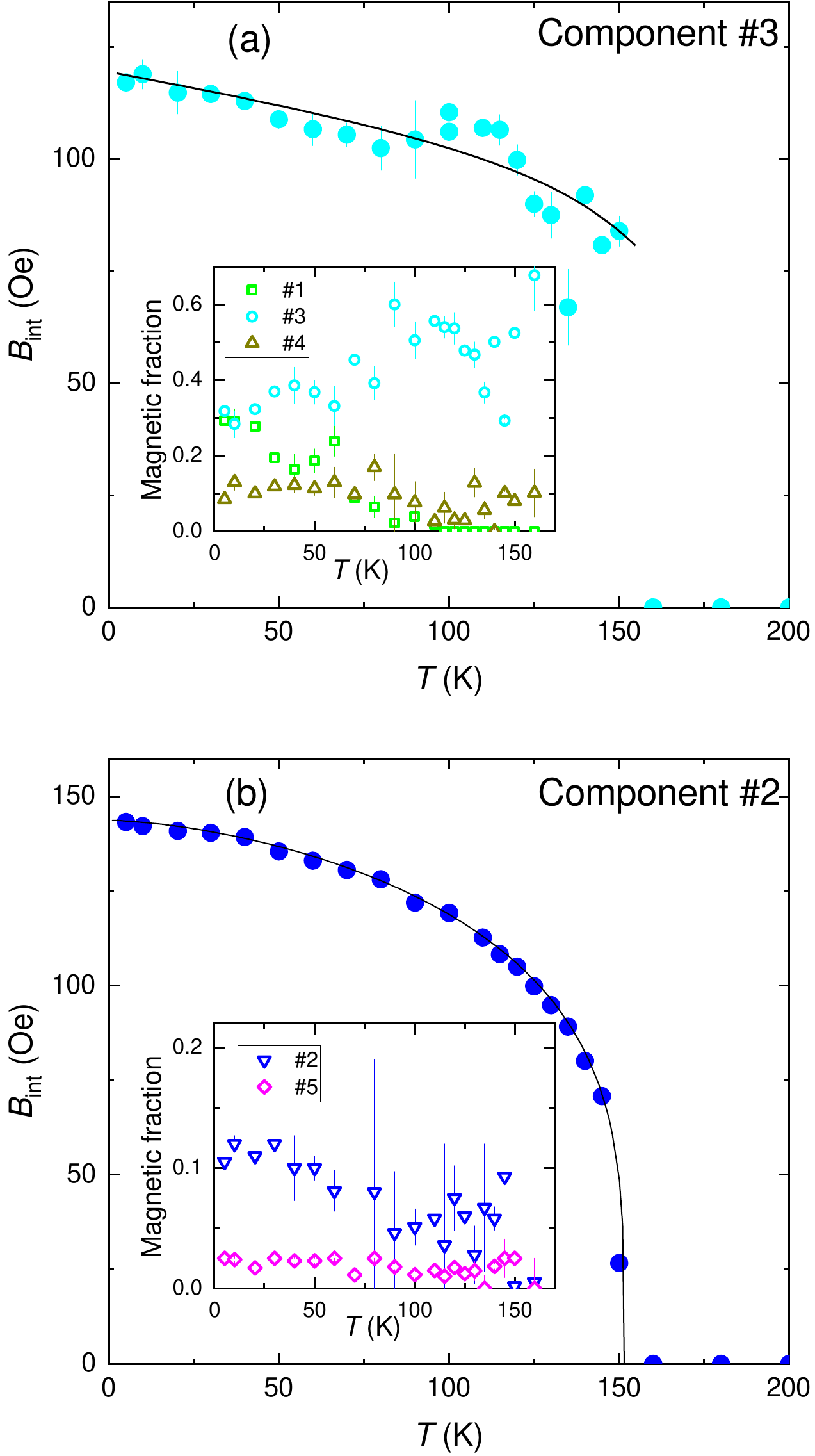}
\caption{(a) Temperature dependence of the internal field components attributed to the 1313 phase. The solid line is a guide to the eye. The inset shows the temperature dependence of the corresponding magnetic fractions. (b) Temperature dependence of internal fields associated with the 2222 phase. The solid line is a power-law fit from Ref.~\onlinecite{Khasanov_NatPhys_2025}. The inset displays the temperature evolution of the magnetic volume fractions.}
\label{fig:Internal-Fields}
\end{figure}

Figure~\ref{fig:Internal-Fields} presents the temperature evolution of the internal magnetic fields. To reduce the number of fit parameters, the internal fields with Lorentzian lineshapes ($B_{\rm int,2}$ and $B_{\rm int,5}$) and those with Overhauser-type lineshapes ($B_{\rm int,1}$, $B_{\rm int,3}$, and $B_{\rm int,4}$) were grouped, with each group assumed to scale proportionally with temperature: $B_{\rm int,2}(T) \propto B_{\rm int,5}(T)$ and $B_{\rm int,1}(T) \propto B_{\rm int,3}(T) \propto B_{\rm int,4}(T)$. The proportionality constants were treated as temperature-independent parameters during fitting. The internal fields of the components with the largest volume fractions -- $B_{\rm int,2}$ for the Lorentzian group and $B_{\rm int,3}$ for the Overhauser group -- were taken as reference components, and the scaling of the remaining internal fields was performed relative to these.

Figure~\ref{fig:Internal-Fields} shows that the temperature dependencies of the internal fields differ significantly between the `Overhauser'  and `Lorenzian' group of signals. The Overhauser-type components exhibit a sharp onset, with the internal field rising abruptly from zero to approximately 80~mT at $T \simeq 150$~K. This behavior is indicative of a first-order-like magnetic transition for the phase associated with the asymmetric (Overhauser-type) field distribution.
In contrast, the Lorentzian components display a gradual decrease in $B_{\rm int}$ with increasing temperature, vanishing near 150~K. This smooth evolution is characteristic of a second-order magnetic transition.
The insets in Fig.~\ref{fig:Internal-Fields} show the temperature dependence of the corresponding volume fractions for each individual component within the group.

\section{Discussion}

\subsection{Assignment of $\mu$SR Responses to Structural Phases}

At the beginning of the discussion, it is important to note that the sample studied in this work -- a mixture of small single crystals -- contains multiple magnetic and nonmagnetic phases (see Secs.~\ref{sec:optical-characterzation} and \ref{subsec:WTF}). Powder x-ray diffraction on the crushed boule used for the $\mu$SR experiments reveals that the material consists predominantly of the 1313 and 2222 phases (together accounting for approximately 70~wt\%), along with significant contributions from La$_2$NiO$_4$ (about 13~wt\%) and La$_4$Ni$_3$O$_{10}$ (about 21~wt\%)  Ruddlesden-Popper phases, see the Supplemental Information.

Assigning the $\mu$SR magnetic components to specific structural phases requires knowledge of the characteristic magnetic responses of the various RP nickelate members. By combining the $\mu$SR data with the literature results reported in Refs.~\onlinecite{Khasanov_NatPhys_2025, Chen_PRL_2024, Khasanov_LaPr327_Arxiv_2025}, the magnetic transition near 150~K ({\it i.e.}, the second magnetic phase occupying $\simeq 15$\% of the sample volume, Sec.~\ref{subsec:WTF}) is attributed to regions of the 2222–La$_3$Ni$_2$O$_7$ phase. This assignment is supported by the following observations:
\\
    (i) The transition temperature of this component, $T_{\rm m,2} \simeq 150$~K at ambient pressure, and its increase under pressure at a rate of ${\rm d}T_{\rm m,2}/{\rm d}p \simeq 2.5$~K/GPa, Fig.~\ref{fig:pressure_Tm}~(b), are both consistent with previously reported results for 2222 structural modification of La$_3$Ni$_2$O$_7$ \cite{Khasanov_NatPhys_2025, Chen_PRL_2024, Khasanov_LaPr327_Arxiv_2025, Plokhikh_Arxiv_2025, Chen_arxiv_2024, Khasanov_OIE_la327-4310_Arxiv_2025}.
\\
    (ii) The ZF-$\mu$SR response of this phase is characterized by two symmetric internal field components with low-temperature values of $B_{\rm int,4}(5~\mathrm{K}) \simeq 145$~mT and $B_{\rm int,5} \simeq 9$~mT, Figs.~\ref{fig:ZF-signals}~(d) and (e), in agreement with 2222 phase response reported in Refs.~\onlinecite{Khasanov_NatPhys_2025, Khasanov_OIE_la327-4310_Arxiv_2025}.
\\
    (iii) The Lorentzian line shapes of these components, Figs.~\ref{fig:ZF-signals}~(d) and (e), indicate commensurate magnetic order, in agreement with previous observations for the 2222 phase \cite{Khasanov_NatPhys_2025, Chen_PRL_2024, Khasanov_LaPr327_Arxiv_2025, Plokhikh_Arxiv_2025, Chen_arxiv_2024, Khasanov_OIE_la327-4310_Arxiv_2025}.
\\
    (iv) The temperature evolution of the internal field $B_{\rm int,2}$, shown in Fig.~\ref{fig:Internal-Fields}(b), agrees well with the behavior reported in Refs.~\onlinecite{Khasanov_NatPhys_2025, Khasanov_LaPr327_Arxiv_2025, Plokhikh_Arxiv_2025, Khasanov_OIE_la327-4310_Arxiv_2025}. Notably, the solid line in Fig.~\ref{fig:Internal-Fields}(b) represents the power-law fit to $B_{\rm int}(T)$ obtained in Ref.~\onlinecite{Khasanov_NatPhys_2025} for a phase-pure 2222-La$_3$Ni$_2$O$_7$ sample.

Taking into account that the 1313 and 2222 phases of La$_3$Ni$_2$O$_7$ together occupy approximately 70\% of the sample volume (see Supplemental Information) and noting that the 2222 phase occupies $\simeq 15\%$ of the sample volume (see discussion above), it can be inferred that roughly $\sim 55\%$ of the total $\mu$SR response is contributed by the 1313 structural modification of La$_3$Ni$_2$O$_7$. This estimate is consistent with the 55--60\% fraction of the first magnetic phase obtained from the WTF-$\mu$SR analysis (Sec.~\ref{subsec:WTF}), as well as with the sum of the volume fractions of the incommensurate components \#1, \#3, and \#4 obtained in the ZF-$\mu$SR measurements (Sec.~\ref{subsec:ambient-pressure-ZF}).

The remaining contributions observed in $\mu$SR experiments  can be assigned to impurity phases,
namely La$_2$NiO$_4$ (5--10\% from WTF-$\mu$SR and $\simeq 13$~wt\% from x-ray) and
La$_4$Ni$_3$O$_{10}$ (approximately 20\% from WTF-$\mu$SR and $\simeq 21$~wt\% from x-ray).
Previous work on hole-doped (La,Sr)$_2$NiO$_4$ has demonstrated the emergence of spin-density-wave (SDW)
order near 200~K \cite{Anissimova_2014, Buttrey_JSSCh_1986}. A similar effect may occur in oxygen-annealed La$_2$NiO$_4$,
where excess oxygen introduces additional holes in much the same way as Sr doping.

Strong oxygen disorder in monolayer and bilayer Ruddlesden--Popper phases is generally known to suppress long-range
magnetic order \cite{Chen_arxiv_2024, Tranquada_PRL_1993, Tranquada_PRL_1994}. While the magnetic properties of La$_4$Ni$_3$O$_{10}$ under comparable disorder
conditions remain insufficiently studied, the volumetric fraction of this impurity phase is consistent
with the nonmagnetic component observed in the $\mu$SR data.

At the end of this section, we note the excellent agreement between the pressure-dependent magnetic response of the 2222-La$_3$Ni$_2$O$_7$ regions identified in the present crystals and that reported previously for phase-pure powder samples~\cite{Khasanov_NatPhys_2025}. This agreement demonstrates a high reproducibility of the magnetic behavior under hydrostatic conditions. It further confirms that the piston–cylinder pressure cell provides an essentially isotropic compression and indicates that the magnetic response of the 2222 phase is robust with respect to sample form (single crystals versus powders). Moreover, the close correspondence suggests either a weak sensitivity of the 2222 phase to oxygen stoichiometry or that both studies probe samples with similarly optimal oxygen content.

\subsection{Magnetic response of 1313-La$_3$Ni$_2$O$_7$}

Based on the phase assignment of the $\mu$SR signals established above, the magnetic response of
the alternating monolayer--trilayer La$_3$Ni$_2$O$_7$ can be
summarized as follows:\\
(i) The 1313-La$_3$Ni$_2$O$_7$ phase undergoes static magnetic ordering.
The midpoint of the magnetic transition, defined as the temperature at which approximately half
of the 1313 phase becomes magnetically ordered, is $\simeq 123$~K.
At the same time, internal magnetic fields at the muon stopping sites appear abruptly at
$\simeq 150$~K, Figs.~\ref{fig:WTF_nonmagnetic-fraction}, \ref{fig:pressure_WTF}(a), and \ref{fig:Internal-Fields}(a).\\
(ii) The sudden emergence of sizeable internal fields at the transition onset indicates a
first-order-like character of the magnetic transition.
Such behavior is consistent with a scenario in which spin-density-wave order develops
together with, and remains strongly coupled to, a charge-density-wave instability.
Similar intertwined spin--charge ordering was reported for trilayer Ruddlesden--Popper
nickelates such as La$_4$Ni$_3$O$_{10}$ and Pr$_4$Ni$_3$O$_{10}$
\cite{Zhang_NatCom_2020, Samarakoon_PRX_2023}.
Within a Landau framework, this behavior can be understood as arising near a first-order region
of the phase diagram where SDW and CDW order parameters are strongly coupled, as discussed
%by Norman
in Ref.~\onlinecite{Norman_PRB_2025}.\\
(iii) Upon further cooling, a second magnetic anomaly is observed near $T \simeq 100$~K.
This transition is manifested by a qualitative change in the internal field distribution,
which evolves from a single-peak structure to a three-peak structure.
An analogous evolution was observed in 3333-La$_4$Ni$_3$O$_{10}$ and attributed to a
spin-reorientation transition occurring at low temperatures
\cite{Khasanov_La4310_Arxiv_2025, Cao_arxiv_2025}.
Consistent with this interpretation, magnetization measurements on single crystals dominated by
the 1313 phase show pronounced anomalies near both 120~K and 100~K in the
temperature derivative of the susceptibility measured along the $c$ axis,
Fig.~\ref{fig:Polarised-microscopy}(d).\\
(iv) The zero-field $\mu$SR spectra provide further insight into the nature of the magnetic order.
The time spectra cannot be described by simple cosine-like oscillations characteristic of
commensurate magnetism.
Instead, the extracted internal field distributions exhibit extended low-field tails,
Figs.~\ref{fig:chi2_5peak-3peak}(b) and (e), a hallmark of incommensurate magnetic order.
Such behavior is well captured by an Overhauser-type field distribution in the frequency domain
and a zeroth-order Bessel function $J_0$ in the time domain
\cite{Overhauser_JPhysChemSolids_1960, Schenck_PRB_2001, Amato_PRB_2014, Khasanov_PRB_MnP_2016}.\\
(v) Hydrostatic pressure further differentiates the magnetic response of the 1313 polymorph from that
of the bilayer and trilayer systems, Fig.~\ref{fig:scetch}~(a).
The SDW transition temperature decreases linearly with pressure at a rate
${\rm d}T_{\rm SDW}/{\rm d}p = -3.9(8)$~K/GPa.
This negative pressure coefficient is opposite in sign to that observed in
2222-La$_3$Ni$_2$O$_7$, where $T_{\rm SDW}$ increases with pressure with the rate ${\rm d}T_{\rm SDW}/{\rm d}p \simeq 2.5$~K/GPa, but is qualitatively similar
to the behavior of the trilayer 3333-La$_4$Ni$_3$O$_{10}$ compound,
where a much larger suppression rate of ${\rm d}T_{\rm SDW}/{\rm d}p \simeq -13$~K/GPa is reported \cite{Khasanov_La4310_Arxiv_2025}.

Taken together, these observations demonstrate that the magnetic response of the 1313-La$_3$Ni$_2$O$_7$ phase more closely resembles that of the trilayer nickelate 3333-La$_4$Ni$_3$O$_{10}$ than that of the bilayer 2222-La$_3$Ni$_2$O$_7$ system. This similarity can be naturally attributed to the presence of a trilayer NiO$_2$ building block within the 1313 structure. The intervening monolayer likely acts as a spacer that modifies the effective interlayer coupling and electronic dimensionality. In addition, possible charge transfer between the monolayer and trilayer units may further influence the formation and coupling of spin- and charge-density-wave orders~\cite{La_Bollita_PRB_2025, Lechermann_PRM_2024}.

\subsection{Comparison of magnetic responses in 2222, 1313, and 3333 nickelates}

\begin{figure}[htb]
\includegraphics[width=0.95\linewidth]{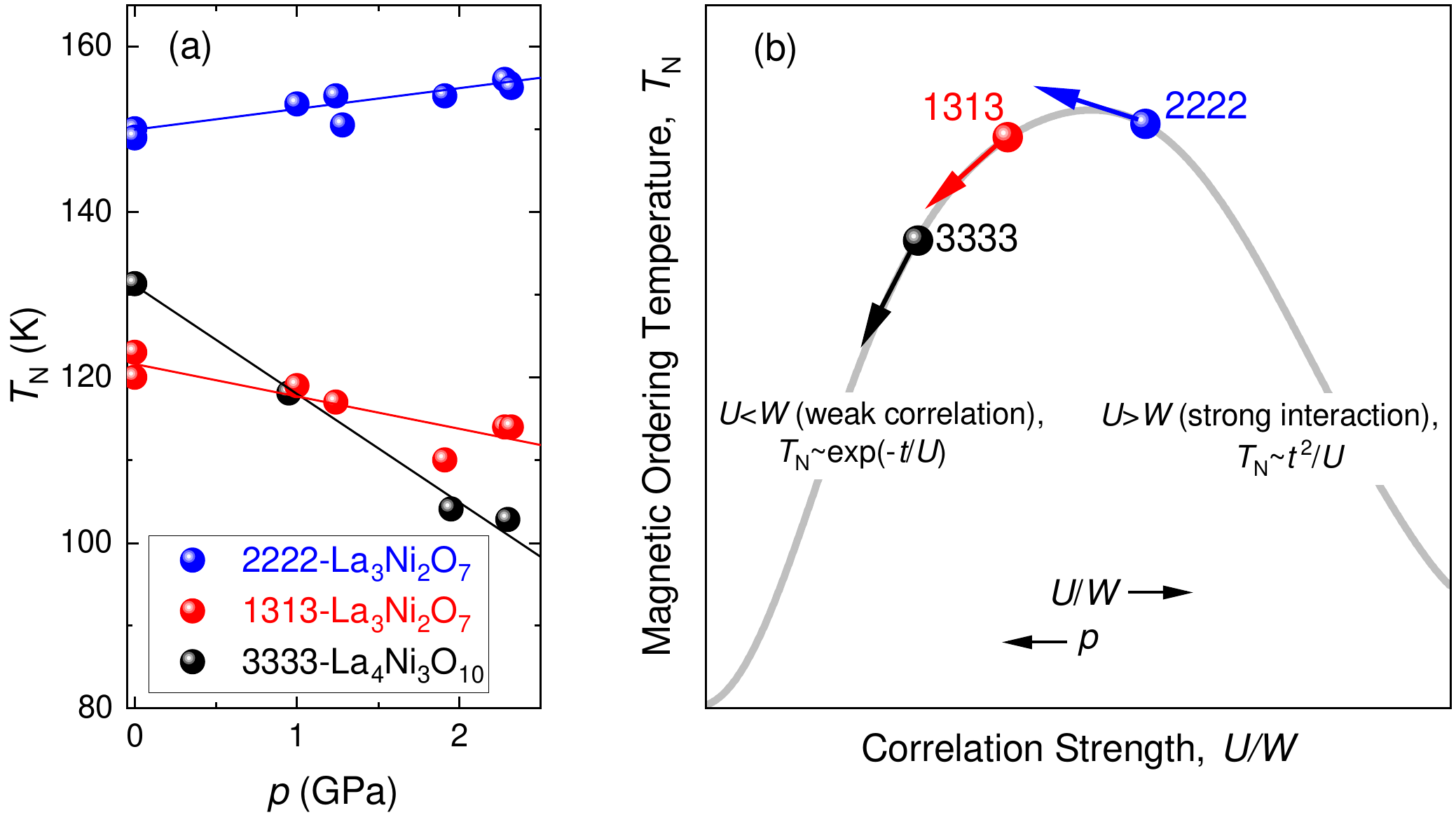}
\caption{
(a) Pressure dependence of the magnetic ordering temperature $T_{\rm N}$ for
2222-La$_3$Ni$_2$O$_7$ (present data together with Ref.~\onlinecite{Khasanov_NatPhys_2025}),
1313-La$_3$Ni$_2$O$_7$ (present study), and
3333-La$_4$Ni$_3$O$_{10}$ (Ref.~\onlinecite{Khasanov_La4310_Arxiv_2025}).
Solid lines represent linear fits with slopes
${\rm d}T_{\rm N}/{\rm d}p \simeq +2.5$~K/GPa for 2222-La$_3$Ni$_2$O$_7$,
${\rm d}T_{\rm N}/{\rm d}p \simeq -3.9$~K/GPa for 1313-La$_3$Ni$_2$O$_7$, and
${\rm d}T_{\rm N}/{\rm d}p \simeq -13$~K/GPa for 3333-La$_4$Ni$_3$O$_{10}$.
(b) Schematic illustration of the magnetic ordering temperature as a function of the
effective electronic correlation strength $U/W$.
Hydrostatic pressure reduces $U/W$, leading to an increase (decrease) of the ordering
temperature for strongly (weakly) correlated electrons.
The relative positions of 2222-La$_3$Ni$_2$O$_7$, 1313-La$_3$Ni$_2$O$_7$, and
3333-La$_4$Ni$_3$O$_{10}$ on the curve reflect their experimentally observed pressure
coefficients as obtained from $T_{\rm N}$ {\it vs.} $p$ shown in panel (a).
}
\label{fig:scetch}
\end{figure}

Figure~\ref{fig:scetch}(b) provides a schematic representation of the dependence of the magnetic
ordering temperature $T_{\rm N}$ on the electronic correlation strength, defined as the ratio $U/W$
of the on-site Coulomb repulsion $U$ to the electronic bandwidth $W$.

In the strongly correlated regime [$U>W$, right-hand side of Fig.~\ref{fig:scetch}(b)], magnetic order
can be described within a local-moment picture. In this case, external pressure enhances the
magnetic exchange interactions,
$J \propto t^{2}/U$ ($t$ is the electronic hopping amplitude).
Since pressure increases $t$, the exchange interaction $J$ is strengthened, leading to an
increase of the magnetic ordering temperature $T_{\rm N} \propto J$.
The resulting spin structure is generally commensurate, provided that the lattice geometry
is not frustrated and that the exchange interactions are dominated by nearest-neighbor couplings.

By contrast, in a less correlated and more itinerant regime [$U<W$, left-hand side of
Fig.~\ref{fig:scetch}(b)], magnetic order is better described as a spin-density-wave
instability driven by Fermi-surface nesting. Such SDW order is generically incommensurate, and
the corresponding ordering temperature is expected to decrease under pressure.
This suppression arises because pressure enhances the hopping amplitude $t$ and, consequently,
the bandwidth $W$, thereby reducing the density of states at the Fermi level,
\(\rho(E_{\rm F}) \propto 1/W\).

Similar considerations apply to the relationship between spin- and charge-density-wave instabilities. In an itinerant picture, Fermi-surface nesting can simultaneously support both SDW and CDW channels, allowing them to cooperate and potentially resulting in a single, first-order transition. In contrast, in a more localized regime, spin and charge degrees of freedom tend to decouple, making separate transitions more natural, as often observed in doped Mott systems.

Applying these general arguments to the Ruddlesden–Popper nickelates, it is inferred that the 2222 phase is more strongly correlated than both the 1313 and 3333 structural phases.
%This conclusion is consistent with recent optical studies, which %suggest a systematic reduction of electronic correlations when %progressing from the 2222 to the 3333 RP nickelate %families~\cite{Liu_PRB_2025}.
Such a trend is consistent with recent optical studies~\cite{Liu_PRB_2025} and is physically plausible, as the trilayer 3333 compound is closer to an isotropic three-dimensional metallic system %, as reflected in
due to its pronounced interlayer hybridization and hence more itinerant electronic character. The 1313 phase likewise exhibits more band-like behavior than the 2222 phase, as inferred from its negative pressure dependence of $T_{\rm N}$ and its tendency toward incommensurate magnetic order.
The role of the monolayer component in the 1313 structure remains an open question, particularly with respect to its intrinsic correlation strength and the extent to which it may be screened by adjacent multilayer units. Addressing this issue will require future layer-resolved experimental probes.

Taken together, the pressure dependence of the magnetic ordering temperature, the (in)commensurability of the spin structure, and the order of the transition consistently indicate that Ni $3d$ electrons are more strongly correlated in the 2222 phase and progressively more itinerant in the 1313 and 3333 phases. This hierarchy of correlation strength further suggests that the higher superconducting transition temperature observed in the 2222 system may be intimately connected to its more strongly correlated electronic nature.

\section{Conclusions}

To conclude, the magnetic properties of the 1313 structural modification of La$_3$Ni$_2$O$_7$ were investigated using muon-spin rotation/relaxation under both ambient and hydrostatic pressure conditions. The monolayer-trilayer phase of La$_3$Ni$_2$O$_7$ exhibits static magnetic order with an onset temperature
$\simeq 150$~K and a midpoint transition temperature
$\simeq 123$~K.
The abrupt appearance of internal magnetic fields at the transition, together with the
Overhauser-like shape of internal field lines, demonstrates that the magnetic order is
first-order-like in character and incommensurate.

Upon application of hydrostatic pressure, the SDW transition temperature is progressively
suppressed at a rate of ${\rm d}T_{\rm SDW}/{\rm d}p = -3.9(8)$~K/GPa.
This pressure evolution closely resembles that of the trilayer nickelate La$_4$Ni$_3$O$_{10}$,
but is opposite to the behavior observed in the bilayer
La$_3$Ni$_2$O$_7$ system.
These results indicate that the magnetic properties of the 1313 La$_3$Ni$_2$O$_7$ are governed primarily
by its trilayer building block, with the intervening monolayer modifying interlayer coupling and
dimensionality.

Within a phenomenological framework developed to compare different Ruddlesden--Popper
nickelates, the combined evolution of the magnetic ordering temperature, the
(in)commensurability of the spin structure, and the character of the magnetic transition
reveals a systematic hierarchy of electronic correlation strength.
The results indicate that Ni $3d$ electrons are most strongly correlated in the bilayer
2222 phase and become progressively more itinerant in the monolayer-trilayer 1313 and the trilayer 3333 phases.
This hierarchy provides a natural explanation for the distinct pressure responses of the SDW
order and suggests that the higher superconducting transition temperature observed in the
2222 system may be intimately connected to its more strongly correlated electronic nature.

Our findings establish monolayer-trilayer La$_3$Ni$_2$O$_7$ as a structural and magnetic bridge between the
bilayer and trilayer Ruddlesden--Popper nickelates.
By linking structural motifs to the evolution of density-wave order under pressure, this work
provides important insight into the competing electronic instabilities in layered nickelates and
sets the stage for understanding their relationship to high-pressure superconductivity.

\section*{ACKNOWLEDGEMENTS}
R.K. acknowledges helpful discussions with Guang-Ming Zhang. Z.G. acknowledges support from the Swiss National Science Foundation (SNSF) through SNSF Starting Grant (No.~TMSGI2${\_}$211750). Dariusz Gawryluk acknowledges support from the Swiss National Science Foundation (SNSF) through Grant No.~200021E\_238113?

\section{Supplementary Material}

A large quantity of small crystals was polished and examined under polarized light. While the majority of the sample consists of the 1313-La$_3$Ni$_2$O$_7$ phase mixed with the polymorph 2222-La$_3$Ni$_2$O$_7$, several regions exhibit mixed Raman spectra, as illustrated by the blue line in Fig.~\ref{boule}(a). In only one location was a clean La$_2$NiO$_4$ Raman signature detected, and in a separate selected region, La$_4$Ni$_3$O$_{10}$ was identified. This behavior was typical across most single crystals. However, in the bulk boule, a larger concentration of impurity phases was observed.

To assess the phase composition, all remaining pieces of the boule—excluding selected single crystals reserved for specific experiments—were crushed into powder and analyzed using powder X-ray diffraction (XRD). The results, shown in Fig.~\ref{boule}(b), reveal significant contributions from adjacent Ruddlesden–Popper (RP) phases, notably the $n = 1$ phase La$_2$NiO$_4$ (12.7 wt\%) and the $n = 3$ phase La$_4$Ni$_3$O$_{10}$ (20.8 wt\%). These impurity phases account for the $\sim$10\% magnetic fraction exhibiting a large N\'eel temperature (T$_N$), as well as the $\sim$20\% nonmagnetic fraction detected via muon spin rotation ($\mu$SR).

The occurrence of these neighboring RP phases may be attributed either to the intrinsic temperature gradient of the vertical mirror optical floating zone  furnace—which is known to influence phase stability within the extremely narrow temperature window for the $n=2$ La$_3$Ni$_2$O$_7$ phase~\cite{Zinkevich2004}—or to the internal pressure gradient caused by limited oxygen diffusion within the boule, leading to variable nickel oxidation states. Similar behavior has been observed in perovskite nickelates, where oxygen nonstoichiometry results in defect structures in LaNiO$_3$~\cite{Zhou2020} and in the formation of RP inclusions in PrNiO$_3$ or doped LaNiO$_3$~\cite{Puphal2023}.

To clarify the stability range of the $n=2$ phase, we conducted growth experiments under varying oxygen partial pressures using stoichiometric feed and seed rods of La$_3$Ni$_2$O$_7$. Based on earlier studies,\cite{Zhang2020} the general phase window lies between 10 and 20 bar. At an oxygen pressure of 10 bar, a mixture of La$_2$NiO$_4$ and La$_4$Ni$_3$O$_{10}$ forms, with only a small fraction of La$_3$Ni$_2$O$_7$. Gradually increasing the pressure to 15 bar results in a steady increase in the La$_4$Ni$_3$O$_{10}$ content, while the La$_3$Ni$_2$O$_7$ fraction remains low. Notably, at exactly 15 bar, a significant increase in La$_3$Ni$_2$O$_7$ content is observed. However, due to the compositional drift caused by preferential crystallization of off-stoichiometric La$_4$Ni$_3$O$_{10}$, the liquid composition deviates from the initial stoichiometry.

Although the floating zone method compensates for this effect by continuously feeding the correct stoichiometry, growth dynamics would significantly differ when starting directly at 15 bar. A slight increase in oxygen pressure beyond this point rapidly suppresses the formation of La$_3$Ni$_2$O$_7$. Therefore, we conclude that the phase stability region for La$_3$Ni$_2$O$_7$ is extremely narrow and centered around 15 bar oxygen partial pressure.

This result explains the frequent appearance of impurity phases in large boules grown at 15 bar: the imposed external gas pressure is only realized near the surface, while the effective internal oxygen partial pressure in the melt is governed by limited diffusion. Diffusion constraints are particularly significant at high pressure and are further exacerbated by the vertical single-lamp design of the furnace~\cite{Phelan2018, Dossa2022}.

\setcounter{figure}{0}
\renewcommand{\figurename}{Fig.}
\renewcommand{\thefigure}{S\arabic{figure}}

\begin{figure}[t]
		\includegraphics[width=1\columnwidth]{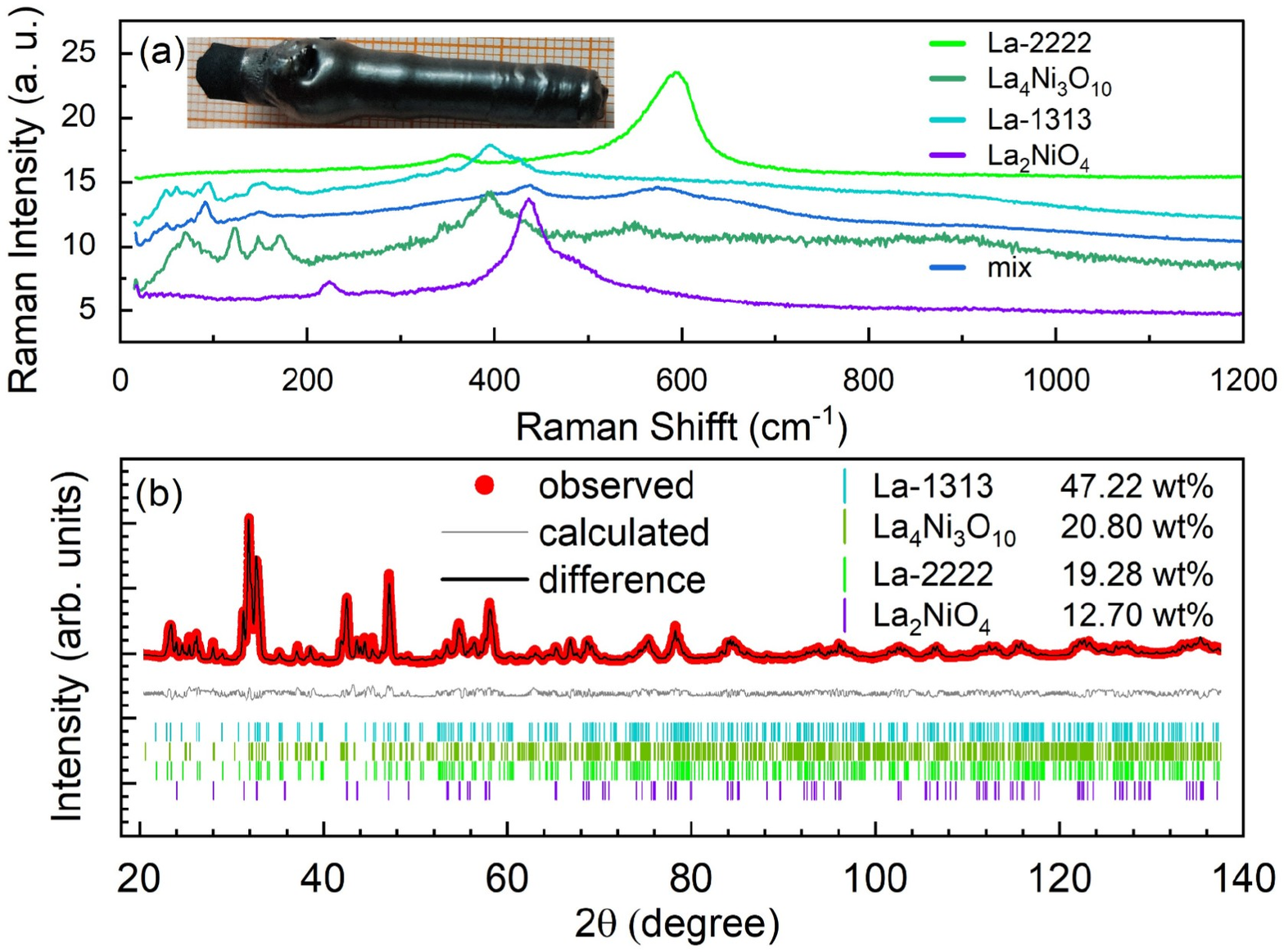}
	\caption{(a) Raman spectra on polished crystals of the La$_3$Ni$_2$O$_7$ boule grown at constant 15 bar pressure shown on the inset. The Raman spectra correspond RP phases 2222 (green), La$_4$Ni$_3$O$_{10}$ (dark green), 1313 (blue), La$_2$NiO$_{4}$ (purple) and a mixed Raman spectra (dark blue). (b) Powder XRD of a large quantity of crushed crystals with an underlying Rietveld refinement. The solid black line corresponds to the calculated intensity from the Rietveld refinement, the solid gray line is the difference between the experimental and calculated intensities, and the vertical blue/green/purple bars are the calculated Bragg peak positions. The corresponding weight fractions are given in the figure next to the phases.}
	\label{boule}
\end{figure}

\begin{figure}[t]
		\includegraphics[width=1\columnwidth]{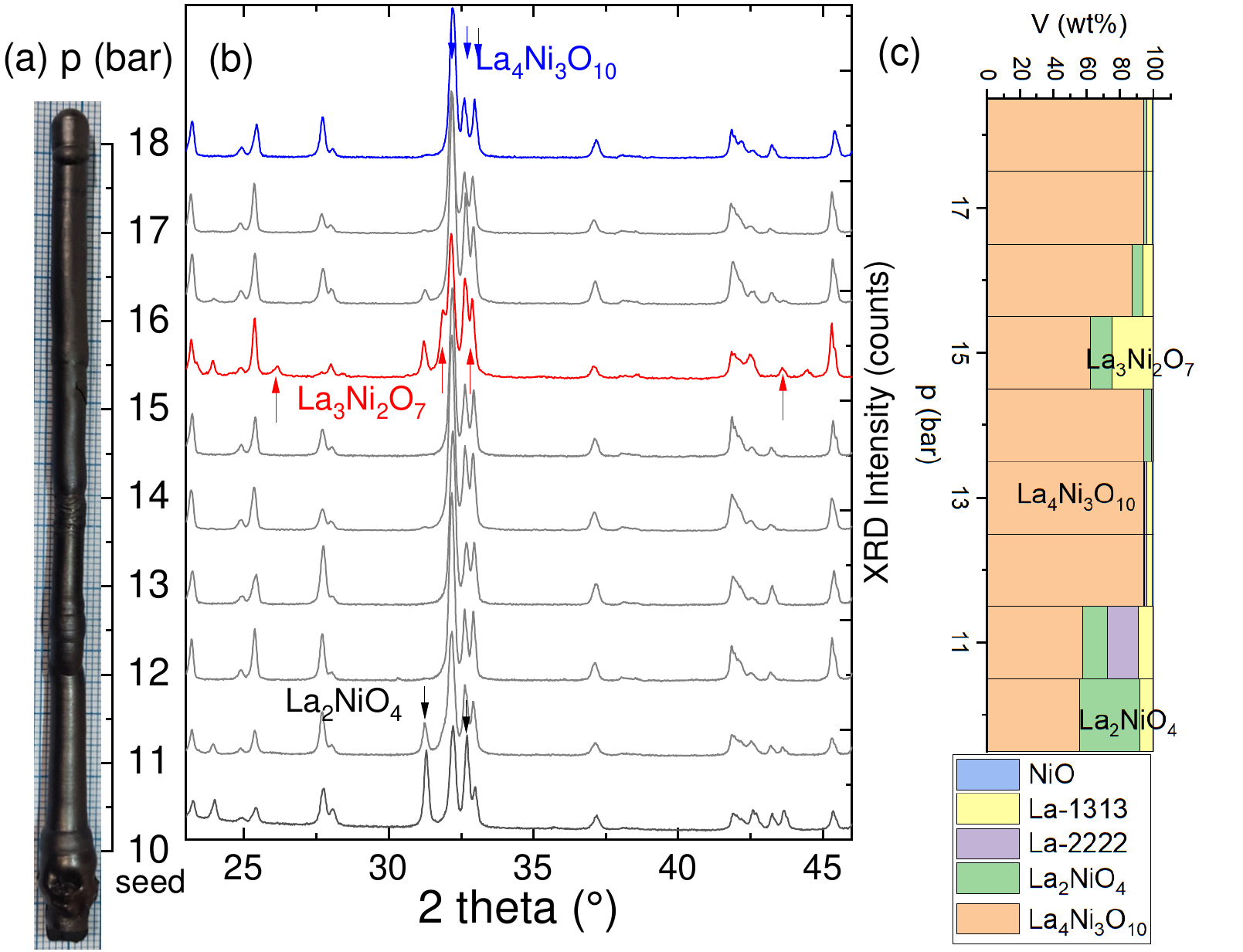}
	\caption{(a) Picture of a large boule of a stoichiometry of
    "La$_3$Ni$_2$O$_7$" gorwn under varying oxygen partial pressures given by the scale on the right. (b) Corresponding XRD patterns of crushed pieces of the different sections of the grown boule, with the main reflexes of the La$_2$NiO$_4$ phase, the La$_3$Ni$_2$O$_7$ phase and the La$_4$Ni$_3$O$_{10}$ phase. (c) Volume estimation of the corresponding phases extracted via Rietveld refinement given in wt\%. }
	\label{gradient}
\end{figure}


\begin{thebibliography}{99}

\bibitem{Sun_Nature_2023} Hualei Sun, Mengwu Huo, Xunwu Hu, Jingyuan Li, Zengjia Liu, Yifeng Han, Lingyun Tang, Zhongquan Mao, Pengtao Yang, Bosen Wang, Jinguang Cheng, Dao-Xin Yao, Guang-Ming Zhang, and Meng Wang,  {\it Signatures of superconductivity near 80~K in a nickelate under high pressure}, Nature {\bf 621}, 493 (2023).\\
    \url{https://doi.org/10.1038/s41586-023-06408-7}

\bibitem{Zhang_NatPhys_2024} Yanan Zhang, Dajun Su, Yanen Huang, Zhaoyang Shan, Hualei Sun, Mengwu Huo, Kaixin Ye, Jiawen Zhang, Zihan Yang, Yongkang Xu, Yi Su, Rui Li, Michael Smidman, Meng Wang, Lin Jiao, and  Huiqiu Yuan, {\it High-temperature superconductivity with zero resistance and strange-metal behaviour in La$_3$Ni$_2$O$_{7-\delta}$}, Nat. Phys. {\bf 20}, 1269 (2024).\\
    \url{https://doi.org/10.1038/s41567-024-02515-y}

\bibitem{Wang_Nature_2024} Ningning Wang, Gang Wang, Xiaoling Shen, Jun Hou, Jun Luo, Xiaoping Ma, Huaixin Yang, Lifen Shi, Jie Dou, Jie Feng, Jie Yang, Yunqing Shi, Zhian Ren, Hanming Ma, Pengtao Yang, Ziyi Liu, Yue Liu, Hua Zhang, Xiaoli Dong, Yuxin Wang, Kun Jiang, Jiangping Hu, Shoko Nagasaki, Kentaro Kitagawa, Stuart Calder, Jiaqiang Yan, Jianping Sun, Bosen Wang, Rui Zhou, Yoshiya Uwatoko, and Jinguang Cheng, {\it Bulk high-temperature superconductivity in pressurized tetragonal La$_2$PrNi$_2$O$_7$}, Nature {\bf 634}, 579 (2024).\\
    \url{https://doi.org/10.1038/s41586-024-07996-8}

\bibitem{Liu_NatCom_2024} Zhe Liu, Mengwu Huo, Jie Li, Qing Li, Yuecong Liu, Yaomin Dai, Xiaoxiang Zhou, Jiahao Hao, Yi Lu, Meng Wang, and Hai-Hu Wen, {\it Electronic correlations and energy gap in the bilayer nickelate La$_3$Ni$_2$O$_7$}, Nat. Commun. {\bf 15}, 7570 (2024).\\
    \url{https://doi.org/10.1038/s41467-024-52001-5}

\bibitem{Zhang_JMST_2024} Mingxin Zhang, Cuiying Pei, Qi Wang, Yi Zhao, Changhua Li, Weizheng Cao, Shihao Zhu, Juefei Wu, and Yanpeng Qi, {\it Effects of pressure and doping on Ruddlesden-Popper phases La$_{n+1}$Ni$_n$O$_{3n+1}$}, J. Mater. Sci. Technol. {\bf 185}, 147 (2024).\\
    \url{https://doi.org/10.1016/j.jmst.2023.11.011}

\bibitem{Li_ChinPhysLet_2024} Yidian Li, Xian Du, Yantao Cao, Cuiying Pei, Mingxin Zhang, Wenxuan Zhao, Kaiyi Zhai, Runzhe Xu, Zhongkai Liu, Zhiwei Li, Jinkui Zhao, Gang Li, Yanpeng Qi, Hanjie Guo, Yulin Chen, and Lexian Yang, {\it Electronic Correlation and Pseudogap-Like Behavior of High-Temperature Superconductor La$_3$Ni$_2$O$_7$}, Chinese Phys. Lett. {\bf 41}, 087402 (2024).\\
    \url{https://doi.org/10.1088/0256-307X/41/8/087402}

\bibitem{Wang_ChinPhysLett_2024} Meng Wang, Hai-Hu Wen, Tao Wu, Dao-Xin Yao, and Tao Xiang, {\it  Normal and superconducting properties of La$_3$Ni$_2$O$_7$},  Chinese Phys. Lett. {\bf 41}, 077402 (2024).\\
    \url{https://doi.org/10.1088/0256-307X/41/7/077402}

\bibitem{Zhou_arxiv_2023} Yazhou Zhou, Jing Guo, Shu Cai, Hualei Sun, Pengyu Wang, Jinyu Zhao, Jinyu Han, Xintian Chen, Yongjin Chen, Qi Wu, Yang Ding, Tao Xiang, Ho-kwang Mao, and Liling Sun, {\it Investigations of key issues on the reproducibility of high Tc superconductivity emerging from compressed La$_3$Ni$_2$O$_7$}, Matter Radiat. Extremes {\bf 10}, 027801 (2025).\\
    \url{https://doi.org/10.1063/5.0247684}

\bibitem{Li_CPL_La4Ni3O10_2024} Q. Li, Y.-J. Zhang, Z.-N. Xiang, Y. Zhang, X. Zhu, and H.-H. Wen, {\it Signature of superconductivity in pressurized La$_4$Ni$_3$O$_{10}$}, Chin. Phys. Lett. {\bf 41}, 017401 (2024).\\
    \url{https://doi.org/10.1088/0256-307X/41/1/017401}

\bibitem{Zhu_Nature_2024} Yinghao Zhu, Di Peng, Enkang Zhang, Bingying Pan, Xu Chen, Lixing Chen, Huifen Ren, Feiyang Liu, Yiqing Hao, Nana Li, Zhenfang Xing, Fujun Lan, Jiyuan Han, Junjie Wang, Donghan Jia, Hongliang Wo, Yiqing Gu, Yimeng Gu, Li Ji, Wenbin Wang, Huiyang Gou, Yao Shen, Tianping Ying, Xiaolong Chen, Wenge Yang, Huibo Cao, Changlin Zheng, Qiaoshi Zeng, Jian-gang Guo, and Jun Zhao, {\it  Superconductivity in pressurized trilayer La$_4$Ni$_3$O$_{10-\delta}$ single crystals}, Nature {\bf 631}, 531 (2024).\\
    \url{https://doi.org/10.1038/s41586-024-07553-3}

\bibitem{Zhang_PRX_2025} Mingxin Zhang, Cuiying Pei, Di Peng, Xian Du, Weixiong Hu, Yantao Cao, Qi Wang, Juefei Wu, Yidian Li, Huanyu Liu, Chenhaoping Wen, Jing Song, Yi Zhao, Changhua Li, Weizheng Cao, Shihao Zhu, Qing Zhang, Na Yu, Peihong Cheng, Lili Zhang, Zhiwei Li, Jinkui Zhao, Yulin Chen, Changqing Jin, Hanjie Guo, Congjun Wu, Fan Yang, Qiaoshi Zeng, Shichao Yan, Lexian Yang, and Yanpeng Qi, {\it Superconductivity in trilayer nickelate La$_4$Ni$_3$O$_{10}$ under pressure}, Phys. Rev. X {\bf 15}, 021005 (2025).\\
    \url{https://doi.org/10.1103/PhysRevX.15.021005}

\bibitem{Puphal_PRL_2024} P. Puphal, P. Reiss, N. Enderlein, Y.-M. Wu, G. Khaliullin, V. Sundaramurthy, T. Priessnitz, M. Knauft, A. Suthar, L. Richter, M. Isobe, P. A. van Aken, H. Takagi, B. Keimer, Y. E. Suyolcu, B. Wehinger, P. Hansmann, and M. Hepting, {\it Unconventional crystal structure of the high-pressure superconductor La$_3$Ni$_2$O$_7$}, Phys. Rev. Lett. {\bf 133}, 146002 (2024).\\
    \url{https://doi.org/10.1103/PhysRevLett.133.146002}

\bibitem{Chen_JACS_2024} X. Chen, J. Zhang, A. S. Thind, S. Sharma, H. LaBollita, G. Peterson, H. Zheng, D. P. Phelan, A. S. Botana, R. F. Klie, and J. F. Mitchell, {\it Polymorphism in the Ruddlesden–Popper nickelate La$_3$Ni$_2$O$_7$: Discovery of a hidden phase with distinctive layer stacking}, J. Am. Chem. Soc. {\bf 146}, 3640 (2024).\\
    \url{https://doi.org/10.1021/jacs.3c14052}

\bibitem{Wnag_InorgChem_2024} H. Wang, L. Chen, A. Rutherford, H. Zhou, and W. Xie, {\it Long-range structural order in a hidden phase of Ruddlesden–Popper bilayer nickelate La$_3$Ni$_2$O$_7$}, Inorg. Chem. {\bf 63}, 5020 (2024).\\
    \url{https://doi.org/10.1021/acs.inorgchem.3c04474}

\bibitem{Abadi_PRL_2025} Sebastien N. Abadi, Ke-Jun Xu, Eder G. Lomeli, Pascal Puphal, Masahiko Isobe, Yong Zhong, Alexei V. Fedorov, Sung-Kwan Mo, Makoto Hashimoto, Dong-Hui Lu, Brian Moritz, Bernhard Keimer, Thomas P. Devereaux, Matthias Hepting, and Zhi-Xun Shen, {\it Electronic structure of the alternating monolayer-trilayer phase of La$_3$Ni$_2$O$_7$}, Phys. Rev. Lett. {\bf 134}, 126001 (2025).\\
    \url{https://doi.org/10.1103/PhysRevLett.134.126001}

\bibitem{Huang_Arxiv_2025} Chaoxin Huang, Jingyuan Li, Xing Huang, Hengyuan Zhang, Deyuan Hu, Mengwu Huo, Xiang Chen, Zhen Chen, Hualei Sun, and Meng Wang, {\it Superconductivity in monolayer-trilayer phase of LaNiO under high pressure}, arXiv:2510.12250.\\
    \url{https://doi.org/10.48550/arXiv.2510.12250}

\bibitem{Li_PRM_2024} F. Li, N. Guo, Q. Zheng, Y. Shen, S. Wang, Q. Cui, C. Liu, S. Wang, X. Tao, G.-M. Zhang, and J. Zhang, {\it Design and synthesis of three-dimensional hybrid Ruddlesden-Popper nickelate single crystals}, Phys. Rev. Mater. {\bf 8}, 053401 (2024).\\
    \url{https://doi.org/10.1103/PhysRevMaterials.8.053401}

\bibitem{Shi_NatPhys_2025} Mengzhu Shi, Di Peng, Kaibao Fan, Zhenfang Xing, Shaohua Yang, Yuzhu Wang, Houpu Li, Rongqi Wu, Mei Du, Binghui Ge, Zhidan Zeng, Qiaoshi Zeng, Jianjun Ying, Tao Wu, and Xianhui Chen, {\it Pressure induced superconductivity in hybrid Ruddlesden-Popper La$_5$Ni$_3$O$_{11}$ single crystals}, Nat. Phys. {\bf 21}, 1780 (2025). \\
    \url{https://doi.org/10.1038/s41567-025-03023-3}

\bibitem{Khasanov_NatPhys_2025} Rustem Khasanov, Thomas J. Hicken, Dariusz J. Gawryluk, Vahid Sazgari, Igor Plokhikh, Lo\"{\i}c Pierre Sorel, Marek Bartkowiak, Steffen B\"{o}tzel, Frank Lechermann, Ilya M. Eremin, Hubertus Luetkens, and Zurab Guguchia, {\it Pressure-enhanced splitting of density wave transitions in La$_3$Ni$_2$O$_{7–\delta}$}, Nat. Phys. {\bf 21}, 430 (2025).\\
    \url{https://doi.org/10.1038/s41567-024-02754-z}

\bibitem{Chen_PRL_2024} Kaiwen Chen, Xiangqi Liu2, Jiachen Jiao, Muyuan Zou, Chengyu Jiang, Xin Li, Yixuan Luo, Qiong Wu,  Ningyuan Zhang, Yanfeng Guo, and Lei Shu, {\it Evidence of Spin Density Waves in La$_3$Ni$_2$O$_{7-\delta}$}, Phys. Rev. Lett. {\bf 132}, 256503 (2024). \\
    \url{https://doi.org/10.1103/PhysRevLett.132.256503}

\bibitem{Chen_arxiv_2024} K. W. Chen, X. Q. Liu, Y. Wang, Z. Y. Zhu, J. C. Jiao, C. Y. Jiang, Y. F. Guo, L. Shu,
    {\it Effect of Pr-doping and oxygen vacancies on spin density wave in La$_3$Ni$_2$O$_{7-\delta}$: A $\mu$SR investigation},
    Phys. Rev. Research {\bf 7}, L032014 (2025). \\
    \url{https://doi.org/10.1103/blkt-x2ps}

\bibitem{Dan_SciBull_2025} Dan Zhao, Yanbing Zhou, Mengwu Huo, Yu Wang, Linpeng Nie, Ye Yang, Jianjun Ying, Meng Wang, Tao Wu, and Xianhui Chen, {\it Pressure-enhanced spin-density-wave transition in double-layer nickelate La$_3$Ni$_2$O$_{7-\delta}$}, Science Bulletin {\bf 70}, 1239 (2025).\\
    \url{https://doi.org/10.1016/j.scib.2025.02.019}

\bibitem{Kakoi_JPSJ_2024} Masataka Kakoi, Takashi Oi, Yujiro Ohshita, Mitsuharu Yashima, Kazuhiko Kuroki, Takeru Kato, Hidefumi Takahashi, Shintaro Ishiwata, Yoshinobu Adachi, Naoyuki Hatada, Tetsuya Uda, and Hidekazu Mukuda, {\it Multiband Metallic Ground State in Multilayered Nickelates La$_3$Ni$_2$O$_7$ and La$_4$Ni$_3$O$_{10}$ Probed by $^{139}$La-NMR at Ambient Pressure}, J. Phys. Soc. Jpn. {\bf 93}, 053702 (2024). \\
    \url{https://doi.org/10.7566/JPSJ.93.053702}

\bibitem{Chen_NatPhys_2024} Xiaoyang Chen, Jaewon Choi, Zhicheng Jiang, Jiong Mei, Kun Jiang, Jie Li, Stefano Agrestini, Mirian Garcia-Fernandez, Xing Huang, Hualei Sun, Dawei Shen, Meng Wang, Jiangping Hu, Yi Lu, Ke-Jin Zhou, and Donglai Feng, {\it Electronic and magnetic excitations in La$_3$Ni$_2$O$_7$}, Nat. Commun. {\bf 15}, 9597 (2024).\\
    \url{https://doi.org/10.1038/s41467-024-53863-5}

\bibitem{Ren_CommPhys_2025} Xiaolin Ren, Ronny Sutarto, Xianxin Wu, Jianfeng Zhang, Hai Huang, Tao Xiang, Jiangping Hu, Riccardo Comin, X. J. Zhou, and Zhihai Zhu, {\it Resolving the Electronic Ground State of La$_3$Ni$_2$O$_{7-\delta}$ Films}, Commun. Phys. {\bf 8}, 52 (2025). \\
    \url{https://doi.org/10.1038/s42005-025-01971-z}

\bibitem{Gupta_NatCom_2025} N. K Gupta, R. Gong, Y. Wu, M. Kang, C. T. Parzyck, B. Z. Gregory, N. Costa, R. Sutarto, S. Sarker, A. Singer, D. G. Schlom, K. M. Shen, and D. G. Hawthorn, {\it Anisotropic Spin Stripe Domains in Bilayer La$_3$Ni$_2$O$_7$}, 	Nat. Commun. {\bf 16}, 6560 (2025). \\
    \url{https://doi.org/10.1038/s41467-025-61653-w}

\bibitem{Luo_CinPhysLett_2025} J. Luo, J. Feng, G. Wang, N. N. Wang, J. Dou, A. F. Fang, J. Yang, J. G. Cheng, Guo-qing Zheng, and R. Zhou, {\it Microscopic evidence of charge- and spin-density waves in La$_3$Ni$_2$O$_7$ revealed by $^{139}$La-NQR}, Chinese Phys. Lett. {\bf 42}, 067402 (2025).\\
    \url{https://doi.org/10.1088/0256-307X/42/6/067402}

\bibitem{Khasanov_La4310_Arxiv_2025} Rustem Khasanov, Thomas J. Hicken, Igor Plokhikh, Vahid Sazgari, Lukas Keller, Vladimir Pomjakushin, Marek Bartkowiak, Szymon Kr\'{o}lak, Micha{\l} J. Winiarski, Jonas A. Krieger, Hubertus Luetkens, Tomasz Klimczuk, Dariusz J. Gawryluk, and Zurab Guguchia, {\it Identical Suppression of Spin and Charge Density Wave Transitions in La$_4$Ni$_3$O$_{10}$ by Pressure}, arXiv:2503.04400.\\
    \url{https://doi.org/10.48550/arXiv.2503.04400}

\bibitem{Zhang_NatCom_2020} Junjie Zhang, D. Phelan, A. S. Botana, Yu-Sheng Chen, Hong Zheng, M. Krogstad, Suyin Grass Wang, Yiming Qi, J. A. Rodriguez-Rivera, R. Osborn, S. Rosenkranz, M. R. Norman, and J. F. Mitchell, {\it Intertwined density waves in a metallic nickelate}, Nat. Commun. {\bf 11}, 6003 (2020).\\
    \url{https://doi.org/10.1038/s41467-020-19836-0}

\bibitem{Cao_arxiv_2025} Yantao Cao, Andi Liu, Bin Wang, Mingxin Zhang, Yanpeng Qi, Thomas J. Hicken, Hubertus Luetkens, Zhendong Fu, Jason S. Gardner, Jinkui Zhao, and Hanjie Guo, {\it Complex spin-density-wave ordering in La$_4$Ni$_3$O$_{10}$}, arXiv:2503.14128.\\
    \url{https://doi.org/10.48550/arXiv.2503.14128}

\bibitem{Khasanov_LaPr327_Arxiv_2025} Rustem Khasanov, Igor Plokhikh, Thomas J. Hicken, Hubertus Luetkens, Dariusz J. Gawryluk, and Zurab Guguchia, {\it Pressure Effect on the Spin Density Wave Transition in La$_2$PrNi$_2$O$_{6.96}$}, Phys. Rev. Research {\bf 7}, L022046 (2025).\\
    \url{https://doi.org/10.1103/PhysRevResearch.7.L022046}

\bibitem{Plokhikh_Arxiv_2025} Igor Plokhikh, Thomas J. Hicken, Lukas Keller, Vladimir Pomjakushin, Samuel H. Moody, Pascale Foury-Leylekian, Jonas J. Krieger, Hubertus Luetkens, Zurab Guguchia, Rustem Khasanov, and Dariusz Jakub Gawryluk, {\it Unraveling Spin Density Wave Order in Layered Nickelates La$_3$Ni$_2$O$_7$ and La$_2$PrNi$_2$O$_7$ via Neutron Diffraction}, arXiv:2503.05287.\\
    \url{https://doi.org/10.48550/arXiv.2503.05287}

\bibitem{Meng_NatCom_2024} Yanghao Meng, Yi Yang, Hualei Sun, Sasa Zhang, Jianlin Luo, Liucheng Chen, Xiaoli Ma, Meng Wang, Fang Hong, Xinbo Wang, and  Xiaohui Yu, {\it Density-wave-like gap evolution in La$_3$Ni$_2$O$_7$ under high pressure revealed by ultrafast optical spectroscopy}, Nat.Commun. {\bf 15}, 10408 (2024).\\
    \url{https://doi.org/10.1038/s41467-024-54518-1}

\bibitem{LiMintago_arxiv_2025} Mingtao Li, Mingxin Zhang, Yiming Wang, Jiayi Guan, Nana Li, Cuiying Pei, N-Diaye Adama, Qingyu Kong, Yanpeng Qi, Wenge Yang, {\it Orbital Signatures of Density Wave Transition in La$_3$Ni$_2$O$_{7-\delta}$ and La$_2$PrNi$_2$O$_{7-\delta}$ RP-Nickelates Probed via in-situ X-ray Absorption Near-edge Spectroscopy}, arXiv:2502.10962. \\
    \url{https://doi.org/10.48550/arXiv.2502.10962}

\bibitem{Au-Yeung_arxiv_2025} Christine C. Au-Yeung, X. Chen, S. Smit, M. Bluschke, V. Zimmermann, M. Michiardi, P. C. Moen, J. Kraan, C. S. B. Pang, C. T. Suen, S. Zhdanovich, M. Zonno, S. Gorovikov, Y. Liu, G. Levy, I. S. Elfimov, M. Berciu, G. A. Sawatzky, J. F. Mitchell, and A. Damascelli, {\it Universal electronic structure of multi-layered nickelates via oxygen-centered planar orbitals}, arXiv:2502.20450v2. \\
    \url{https://doi.org/10.48550/arXiv.2502.20450}


\bibitem{Amato_RSI_2017} A. Amato, H. Luetkens, K. Sedlak, A. Stoykov, R. Scheuermann, M. Elender, A. Raselli, and D. Graf, {\it The new versatile general purpose surface-muon instrument (GPS) based on silicon photomultipliers for $\mu$SR measurements on a continuous-wave beam}, Rev. Sci. Instrum. {\bf 88}, 093301 (2017).\\
    \url{https://doi.org/10.1063/1.4986045}

\bibitem{Khasanov_HPR_2016} R. Khasanov, Z. Guguchia, A. Maisuradze, D. Andreica, M. Elender, A. Raselli, Z. Shermadini, T. Goko, F. Knecht, E. Morenzoni, and A. Amato, {\it High pressure research using muons at the Paul Scherrer Institute}, High Pressure Res. {\bf 36}, 140 (2016).\\
    \url{https://doi.org/10.1080/08957959.2016.1173690}

\bibitem{Khasanov_JAP_2022} Rustem Khasanov, {\it Perspective on muon-spin rotation/relaxation under hydrostatic pressure}, J. Appl. Phys. {\bf 132}, 190903 (2022).\\
    \url{https://doi.org/10.1063/5.0119840}

\bibitem{Amato-Morenzoni_book_2024} Alex Amato and Elvezio Morenzoni, {\it Introduction to Muon Spin Spectroscopy. Applications to Solid State and Material Sciences}, Springer Nature (2024).\\
    \url{https://doi.org/10.1007/978-3-031-44959-8}

\bibitem{Schenk_book_1995} A. Schenck and F. N. Gygax, {\it Magnetic Materials Studied by Muon Spin Rotation Spectroscopy}, in: Handbook of Magnetic Materials, edited by K. H. J. Buschow, Vol. 9, pp. 57–302, Elsevier, Amsterdam (1995).

\bibitem{Yaouanc_book_2011} A. Yaouanc and P. D. de R\'eotier, {\it Muon Spin Rotation, Relaxation, and Resonance}, Oxford University Press (2011).

\bibitem{Blundell_book_2022} S. J. Blundell, R. De Renzi, T. Lancaster, and F. L. Pratt (eds.), {\it Muon Spectroscopy: An Introduction}, Oxford University Press (2022).

\bibitem{Blundell_AnnRev_2025} Stephen J. Blundell, {\it Muon Studies of Superconductors}, Annu. Rev. Condens. Matter Phys. {\bf 16}, 367 (2025). \\
    \url{https://doi.org/10.1146/annurev-conmatphys-032922-095149}

\bibitem{MUSRFIT} A. Suter and B. Wojek, {\it Musrfit: A Free Platform-Independent Framework for $\mu$SR Data Analysis}, Phys. Procedia {\bf 30}, 69 (2012).\\
    \url{https://doi.org/10.1016/j.phpro.2012.04.042}

\bibitem{Sundaramurthy_arxiv_2025} V. Sundaramurthy, A. Suthar, P. Puphal, C. Le, Y. Gu, H. Yilmaz, P. Sosa-Lizama, P. A. van Aken, Y. E. Suyolcu, M. Isobe, X. Wu, M. Minola, B. Keimer, and M. Hepting, {\it Comparative Raman study of Ruddlesden-Popper nickelates and the monolayer-trilayer polymorph}, arXiv:2512.17583v1. \\
\url{https://doi.org/10.48550/arXiv.2512.17583}

\bibitem{Overhauser_JPhysChemSolids_1960} A. W. Overhauser, {\it Mechanism of antiferromagnetism in dilute alloys}, J. Phys. Chem. Solids {\bf 13}, 71 (1960).\\
    \url{https://doi.org/10.1016/0022-3697(60)90128-1}

\bibitem{Schenck_PRB_2001} A. Schenck, D. Andreica, F. N. Gygax, and H. R. Ott, {\it Extreme quantum behavior of positive muons in CeAl$_2$ below 1~K}, Phys. Rev. B {\bf 65}, 024444 (2001).\\
    \url{https://doi.org/10.1103/PhysRevB.65.024444}

\bibitem{Amato_PRB_2014} A. Amato, P. Dalmas de R\'{e}otier, D. Andreica, A. Yaouanc, A. Suter, G. Lapertot, I. M. Pop, E. Morenzoni, P. Bonf\`{a}, F. Bernardini, and R. De Renzi, {\it Understanding the $\mu$SR spectra of MnSi without magnetic polarons}, Phys. Rev. B {\bf 89}, 184425 (2014).\\
    \url{https://doi.org/10.1103/PhysRevB.89.184425}

\bibitem{Khasanov_PRB_MnP_2016} R. Khasanov, A. Amato, P. Bonf\`{a}, Z. Guguchia, H. Luetkens, E. Morenzoni, R. De Renzi, and N. D. Zhigadlo, {\it High-pressure magnetic state of MnP probed by means of muon-spin rotation}, Phys. Rev. B {\bf 93}, 180509(R) (2016).\\
    \url{https://doi.org/10.1103/PhysRevB.93.180509}

\bibitem{Khasanov_OIE_la327-4310_Arxiv_2025} Rustem Khasanov, Vahid Sazgari, Igor Plokhikh, Marisa Medarde, Ekaterina Pomjakushina, Tomasz Klimczuk, Szymon Królak, Michał J. Winiarski, Thomas J. Hicken, Hubertus Luetkens, Zurab Guguchia, and Dariusz J. Gawryluk, {\it Oxygen-isotope effect on the density wave transitions in La$_3$Ni$_2$O$_7$ and La$_4$Ni$_3$O$_{10}$}, arXiv:2504.08290.\\
    \url{https://doi.org/10.48550/arXiv.2504.08290}

\bibitem{Anissimova_2014} S. Anissimova, D. Parshall, G. D. Gu, K. Marty, M. D. Lumsden, S. Chi, J. A. Fernandez-Baca, D. L. Abernathy, D. Lamago, J. M. Tranquada, and D. Reznik, {\it Direct observation of dynamic charge stripes in La$_{2–x}$Sr$_x$NiO$_4$}, Nat. Commun. {\bf 5}, 3467 (2014).\\
    \url{https://doi.org/10.1038/ncomms4467}

\bibitem{Buttrey_JSSCh_1986} D.J. Buttrey, J.M. Honig , C.N.R. Rao, {\it Magnetic properties of quasi-two-dimensional La$_2$NiO$_4$}, J. Solid State Chem. {\bf 64}, 287 (1986). \\
    \url{https://doi.org/10.1016/0022-4596(86)90073-3}

\bibitem{Tranquada_PRL_1993} J. M. Tranquada, D. J. Buttrey, and D. E. Rice, {\it Phase separation, charge-density waves, and magnetism in ${\mathrm{La}}_{2}$${\mathrm{NiO}}_{4+\mathrm{\ensuremath{\delta}}}$ with \ensuremath{\delta}=0.105}, Phys. Rev. Lett. {\bf 70}, 445 (1993). \\
  \url{https://link.aps.org/doi/10.1103/PhysRevLett.70.445}

\bibitem{Tranquada_PRL_1994} J. M. Tranquada, D. J. Buttrey, V. Sachan, and J. E. Lorenzo,   {\it Simultaneous Ordering of Holes and Spins in ${\mathrm{La}}_{2}$Ni${\mathrm{O}}_{4.125}$}, Phys. Rev. Lett. {\bf 73}, 1003, (1994). \\
    \url{https://link.aps.org/doi/10.1103/PhysRevLett.73.1003}

\bibitem{Samarakoon_PRX_2023} A. M. Samarakoon, J. Strempfer, J. Zhang, F. Ye, Y. Qiu, J.-W. Kim, H. Zheng, S. Rosenkranz, M. R. Norman, J. F. Mitchell, and D. Phelan, {\it Bootstrapped dimensional crossover of a spin density wave}, Phys. Rev. X {\bf 13}, 041018 (2023).\\
    \url{https://doi.org/10.1103/PhysRevX.13.041018}

\bibitem{Norman_PRB_2025} M. R. Norman, {\it Landau theory of the density wave transition in trilayer Ruddlesden-Popper nickelates}, Phys. Rev. B {\bf 112}, 075149 (2025). \\
    \url{https://doi.org/10.1103/43qs-x65n}

\bibitem{La_Bollita_PRB_2025} Harrison La Bollita, Soumen Bag, Jesse Kapeghian, and Antia S. Botana, {\it Electronic correlations, layer distinction, and electron doping in the alternating single-layer-trilayer La$_3$Ni$_2$O$_7$ polymorph}, Phys. Rev. B {\bf 110}, 155145 (2024).\\
    \url{https://doi.org/10.1103/PhysRevB.110.155145}

\bibitem{Lechermann_PRM_2024} Frank Lechermann, Steffen B\"{o}tzel, and Ilya M. Eremin, {\it Electronic instability, layer selectivity, and Fermi arcs in La$_3$Ni$_2$O$_7$}, Phys. Rev. Mater. {\bf 8}, 074802 (2024).\\
    \url{https://doi.org/10.1103/PhysRevMaterials.8.074802}

\bibitem{Liu_PRB_2025} Zhe Liu, Jie Li, Mengwu Huo, Bingke Ji, Jiahao Hao, Yaomin Dai, Mengjun Ou, Qing Li, Hualei Sun, Bing Xu, Yi Lu, Meng Wang, and Hai-Hu Wen, {\it Evolution of electronic correlations in the Ruddlesden-Popper nickelates}, Phys. Rev. B 111, L220505 (2025). \\
\url{https://doi.org/10.1103/byl4-zbfv}


\bibitem{Zinkevich2004} M. Zinkevich and F. Aldinger, {\it Thermodynamic analysis of the ternary La–Ni–O system}, J. Alloys Compd. {\bf 375}, 147 (2004).\\
    \url{https://doi.org/10.1016/j.jallcom.2003.11.138}

\bibitem{Zhou2020} Hong Zheng, Bi-Xia Wang, W. Adam Phelan, Junjie Zhang, Yang Ren, M. Krogstad, S. Rosenkranz, R. Osborn, and J. F. Mitchell, {\it Oxygen Inhomogeneity and Reversibility in Single Crystal LaNiO$_{3-\delta}$}, Crystals {\bf 10}, 557 (2020).\\
    \url{https://doi.org/10.3390/cryst10070557}

\bibitem{Puphal2023} P. Puphal, V. Sundaramurthy, V. Zimmermann, K. Küster, U. Starke, M. Isobe, B. Keimer, and M. Hepting, {\it Phase formation in hole- and electron-doped rare-earth nickelate single crystals}, APL Mater. {\bf 11}, 081107 (2023).\\
    \url{https://doi.org/10.1063/5.0160912}

\bibitem{Zhang2020} Junjie Zhang, Hong Zheng, Yu-Sheng Chen, Yang Ren, Masao Yonemura, Ashfia Huq, and J. F. Mitchell, {\it High oxygen pressure floating zone growth and crystal structure of the metallic nickelates $R_4$Ni$_3$O$_{10}$ ($R$ = La, Pr)}, Phys. Rev. Mater. {\bf 4}, 083402 (2020).\\
    \url{https://doi.org/10.1103/PhysRevMaterials.4.083402}

\bibitem{Phelan2018} W. Adam Phelan, Jessica Zahn, Zachary Kennedy, and Tyrel M. McQueen, {\it Pushing boundaries: High pressure, supercritical optical floating zone materials discovery}, J. Solid State Chem. {\bf 266}, 196 (2018).\\
    \url{https://doi.org/10.1016/j.jssc.2018.12.013}

\bibitem{Dossa2022} Scott S. Dossa and Jeffrey J. Derby, {\it Modeling optical floating zone crystal growth in a high-pressure, single-lamp furnace}, J. Cryst. Growth {\bf 589}, 126723 (2022).\\
    \url{https://doi.org/10.1016/j.jcrysgro.2022.126723}

\end{thebibliography}
\end{document}